\documentclass[prb,twocolumn,showpacs,preprintnumbers,amsmath,amssymb,superscriptaddress]{revtex4}
\usepackage{epsf}
\usepackage{graphicx}
\usepackage{bm}% bold math
\usepackage{amsmath}
\usepackage{color}
\usepackage{notes2bib}
\usepackage{epstopdf}

\bibnotesetup{
note-name = , use-sort-key = false }
\begin{document}

\title{A nontrivial crossover in topological Hall effect regimes}

\author{K.~S.~Denisov}
\email{denisokonstantin@gmail.com} \affiliation{Ioffe Institute, St.Petersburg, 194021 Russia}
\affiliation{Lappeenranta University of Technology, FI-53851 Lappeenranta, Finland}
\author{I.~V.~Rozhansky}
\affiliation{Ioffe Institute,
St.Petersburg, 194021 Russia} \affiliation{Lappeenranta University of
Technology, FI-53851 Lappeenranta, Finland}
\affiliation{ITMO University, St.Petersburg, 197101 Russia}
%\affiliation{Institute of Physics, Nanotechnology and Telecommunications, Peter the Great Saint-Petersburg Polytechnic University, 195251 St. Petersburg, Russia}
\author{N.~S.~Averkiev}
\affiliation{Ioffe Institute, St.Petersburg, 194021 Russia}
\author{E.~L\"ahderanta}
\affiliation{Lappeenranta University of Technology, FI-53851 Lappeenranta, Finland}

\begin{abstract}
We propose a new theory of the topological Hall effect (THE) in systems with chiral magnetization vortices such as magnetic skyrmions.
%Our consideration is based on an electron scattering on a single magnetic skyrmion.
We solve the problem of electron scattering on a magnetic skyrmion exactly,
%Calculating the exact electron scattering cross-section on a single magnetic skyrmion %
for an arbitrary strength of exchange interaction and the skyrmion size.
 We report the existence of different regimes of THE and
  resolve the apparent contradiction between the adiabatic Berry phase theoretical approach and the perturbation theory for THE.
%apparent contradictions between them.
%with a nontrivial crossover between them.
%For strong exchange interaction and large skyrmion size there is a pronounced %transverse spin current. On the contrary, in case of a weak exchange coupling and %small skyrmion size a charge Hall current dominates with spin transverse current %being strongly suppressed.
We traced how the topological charge Hall effect transforms into the spin Hall effect upon varying the exchange interaction strength or the skyrmion size.
This transformation has a nontrivial character: it is accompanied by an oscillating behavior of both charge and spin Hall currents.
This hallmark of THE allows one to differentiate the chirality driven contribution to Hall response in the experiments.
%due to skyrmions when treating
%upon various the parameters one could expect nonmonotonical and oscillating pecularities, that
% help, detect, extract

%of the pure charge Hall effect into the spin Hall effect
%We've investigated how the topological charge Hall effect transforms into the spin Hall effect upon varying the exchange interaction strength or the skyrmion size.
%Therefore our work fills the gap between the two limiting cases and resolves apparent contradictions between them; we trace how the topological charge Hall effect transforms into the spin Hall effect upon varying the exchange interaction strength or the skyrmion size.

\end{abstract}


\date{\today}

\maketitle

\section{Introduction}

Anomalous Hall effect (AHE) has been a subject of intensive experimental research and theoretical debates for over a decade \cite{Dyakonov,AHE-Sinova,Sinitsyn}.
%as well as the burning topic of
The AHE is often classified into extrinsic and intrinsic contributions.
The former is due to spin-dependent scattering of charge carriers and thus depends on the scatterers,
 the intrinsic contribution is described in terms of band structure in crystal momentum space and anomalous velocity causing spin separation for electrons moving in an external electric field~\cite{Dyakonov}.
While attributing a particular mechanism to an experimental data often remains a difficult task,
these AHE mechanisms are based on the same physical phenomena in their origin, the spin-orbit interaction.
For a charged particle with a spin moving in an electric field a magnetic field appears in its moving frame which interacts with the particle's spin. The electric field can be either a built-in crystal field or that produced by an impurity or an external field. The spin-orbit interaction underlying AHE couples particle motion with its spin and
directly leads to the spin separation.
The spin separation in its turn results in a transverse charge separation and a finite Hall response
when the carries are spin polarized, usually due to a macroscopic magnetization of the sample~\cite{AHE-Sinova,AbakumovAHE}.

Along with the normal Hall effect and anomalous Hall effect a fundamentally different phenomena has been recently discovered, namely Topological Hall effect (THE)\cite{MnSiAPhase,LeeOnose}.
%The THE appears when the magnetization field has a nonzero spin chirality $\chi_c = {\bf M}_1 \cdot \bigl[{\bf M}_2 \times {\bf M}_3\bigr]$, where ${\bf M}_i$ $(i=1,2,3)$ are three non-coplanar magnetic moment orientations~\cite{Chir_Wilczek}.
%A finite $\chi_c$ behaves as a magnetic field,
%\textcolor{blue}{
The THE appears in systems with non-coplanar ordering of magnetic moments resulting in a non-zero spin chirality of the magnetization field.
Although the spin-orbit interaction is often responsible for appearance of the chiral magnetization fields\cite{Uzdin-1,Wiesendanger-1,rozsa2016complex}, the charge separation arises from exchange interaction of a mobile electron with a non-trivial spatial configuration of magnetic ion spins and thus it is indeed qualitatively different from the AHE.

The THE has been observed experimentally in quite a few systems including 3D pyrochlore lattices \cite{Taguchi_Science,Machida_PRL}, antiferromagnets\cite{surgers2014large,ueland2012controllable}, spin glasses\cite{SpinGlass1,SpinGlass2}, thin films of EuO \cite{EuO},
%where $\chi_c$ is generated on the short length scale due to spin frustrations;
materials with colossal magnetoresistance\cite{Matl_CMR,Jacob_CMR} and in a 2D diluted magnetic semiconductor (Ga,Mn)As\cite{AronzonRozh}.
%where $\chi_c$ is due the thermal chiral fluctuation; or the spectacular manifistation in materials with large scaled chiral continuous magnetic textures,
%(its topological nature introduces the second name for the discussed phenomena: Topological Hall effect),
An impressive manifestation of THE has been found for various thin films containing magnetic skyrmions - vortex-like topologically non-trivial spatially localized configuration of magnetization field \cite{NagaosaNature}, that produces an observable THE response\cite{DiscretHall}. A pronounced THE has been also observed for magnetic skyrmion lattices in MnSi \cite{Muhl_MnSi_Science,MnSiAPhase,THE_Li,Chapman_PRB}, Fe$_x$Co$_{1-x}$Si \cite{Munzer_PRB}, FeGe \cite{FeGe_THE}, arrays of magnetic skyrmions\cite{Zhou-1,Zhou-2} and other artificial states\cite{gilbert2015realization,li2014tailoring}. This makes magnetic skyrmions considered as
%The topological magnetic defects, as magnetic skyrmions,
%are
new promising objects for applications in novel magnetic devices\cite{NagaosaNature,Parkin},
they can be used for racetrack memory\cite{racetrack,liang2015current,tomasello2014strategy,Fert} with THE based read-out \cite{sk_memory,fert2013skyrmions,zhang2014magnetic}.

Up to now there has been no complete theory of THE describing various magnetic materials with metallic type of conductivity.
%\textcolor{blue}{
%The point is that
The existing theories either make use of the adiabatic Berry phase approach valid for the case of a strong exchange
interaction\cite{BrunoDugaev,Tatara_2007,Ye1999}, calculate the spin-dependent scattering perturbatively in the case of a weak exchange strength\cite{prl_skyrmion,Tatara,Onoda_SkyrmNumber}
or use tight-binding simulations\cite{Arab_Papa,Simu_Chir,Simulation}. These theories give contradictive predictions concerning the role of the carrier spin polarization in THE.
In this paper we suggest a universal theoretical approach
%resolving this issue
%conflict and
%unified theory
capable of describing THE for arbitrary strength of the exchange interaction and structure parameters. We attest to the existence of different regimes of THE and describe the transition between charge Hall and spin Hall topological effects, which has previously lacked proper understanding.

\begin{figure*}
	\centering	
	\includegraphics[width=.65\textwidth]{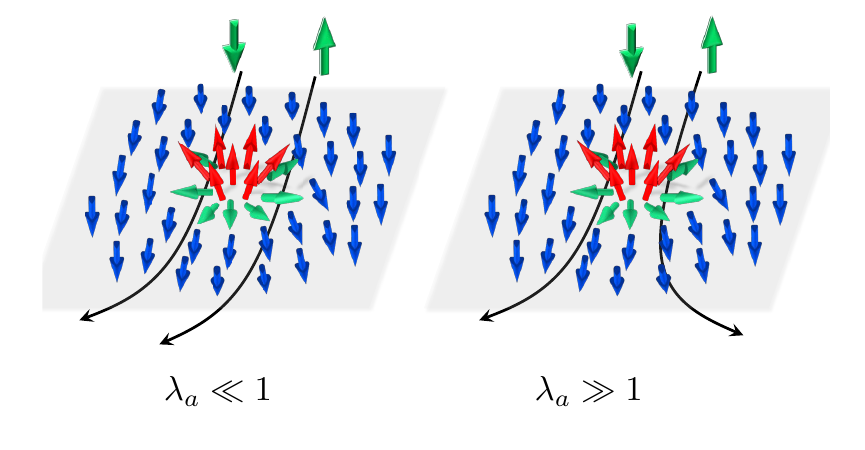}%{fig-cr-2.eps}%{fig-1.pdf}%{fig-1.png}%{3-3.jpg}%{BDS.eps}
	\caption{Electron scattering on a magnetic skyrmion. For a small adiabatic parameter $\lambda_a$ spin-up and spin-down electrons scatter in the same direction resulting in a transverse charge current. For large $\lambda_a$ the scattering in the opposite direction leads to a spin Hall current.}
	\label{figintr1}
\end{figure*}

%Let us consider the limiting regime in depth.
The applicability of various theoretical approaches depends on the adiabatic parameter introduced as $\lambda_a = \omega_{ex} \tau$, where $\hbar \omega_{ex}$ is the spin splitting energy due to a local exchange interaction between an electron and magnetic ions spin texture, $\tau$ is the time of electron's flight through a region of a chiral magnetization field.
%\textcolor{blue}{
In this work we reveal the qualitatively different regimes of THE
with respect to the magnitude of $\lambda_a$.
%We argue that THE exhibits qualitatevely different behaviour with respect to the value %of the adiabatic parameter $\lambda_a$.
%}

Let us consider the limiting cases. The adiabatic regime corresponds to $\lambda_a \gg 1$ (strong exchange interaction and large skyrmion size, typical for strong ferromagnets), at that quantum transitions between spin subleves are suppressed and the carrier spin quantization axis follows the direction of local magnetization. The adiabatic approximation considers the effect of magnetization on the carrier motion via a geometric phase, which the carrier wavefunction acquires while moving through the region with non-zero chirality\cite{Ye1999,BrunoDugaev}. This phase is usually regarded as Berry phase or Pancharatnam phase\cite{berry1984quantal,Lyana-Geller1}. In analogy with Aaronov-Bohm effect this phase can be related to an effective magnetic field. The hallmark of the adiabatic approximation is that this effective magnetic field is opposite for spin-up and spin-down electrons (see Fig.~\ref{figintr1}, right panel); so the polarization of electron gas is essential to produce a transverse charge current response \cite{BrunoDugaev,Tatara_2007}. In this regard it is similar to AHE discussed above where the average polarization of the carriers was needed to produce a transversal charge separation from the spin separation.

The opposite limiting case corresponds to $\lambda_a \ll 1$ (weak exchange interaction and small skyrmion size, typical for spin glasses and diluted magnetic semiconductors). In this case the non-adiabatic perturbation allows quantum transitions between the spin sublevels split by the exchange field, so the appropriate theory should account for the spin-flip scattering  \cite{Onoda_SkyrmNumber,Tatara,prl_skyrmion}. The core prediction of the weak coupling theory is that THE is possible even for spin unpolarized carriers\cite{Tatara,Onoda_SkyrmNumber}. In our previous work\cite{prl_skyrmion} we showed, that this is due to the fact, that while the current of non-polarized carriers flows along the sample the transverse charge separation occurs without
%being accompanied by
spin Hall effect (see Fig.~\ref{figintr1}, left panel). This is in a contrast to the prevailing spin Hall effect at large $\lambda_a$.

In this paper we present a theory covering the whole range of the adiabatic parameter values including the limiting cases of very large or very small $\lambda_a$.
In our approach we calculate an exact scattering cross section of an itinerant carrier on a localized magnetic chiral vortex.
 %with a non-zero chirality.
 We show that the non-zero chirality leads to an asymmetric contribution to the cross section and gives rise to the transverse Hall current. We trace the evolution of this asymmetric contribution with  $\lambda_a$ and describe how the transverse charge
 %pure
  current at  $\lambda_a \ll 1$ transfers into a transverse spin current at $\lambda_a \gg 1$ (see Fig.\ref{figintr1}).
%\textcolor{blue}{
We have found that at $\lambda_a \sim 1$ THE undergoes a nontrivial crossover: both spin and charge Hall currents exhibit oscillatory behavior,
%structure upon variyng  parameters,
which provides a new tool for an experimental detection of THE.
%}
%the crossover the system undergoes in the intermediate range $\lambda_a \sim 1$ has a nontrivial character. Instead of trivial replacement we observe the oscillating structure of both charge and spin currents.
%in the intermediate range of the adiabatic parameter values $\lambda_a \sim 1$ system undergoes a nontrivial resonant crossover.
% advantage simple calculation, has a great powerfull prediction force

\section{General theory}
%We consider a 2D magnetic film with localized chiral magnetization vortices.
We consider an electron in a 2D film scattering on a  magnetic vortex
%vortex.chiral
%single localized vortex
characterized by a non-zero spin chirality. 
%In this section we provide the general theory of electron scattering on a single large scaled structure with
%$\chi_c \neq 0$.   %magnetic skyrmion
We put no restrictions on the adiabatic parameter magnitude.
The electron interacts with the magnetization field by means of exchange interaction.
To extract the pure THE contribution in the following we assume a simple electron band with a quadratic dispersion completely unaffected by spin-orbit interaction. We also neglect dynamics of magnetic centers thus describing the magnetization by a classical vector field ${\bf M\left({\bf r}\right)}$ with the parametrization introduced below.
%The band structure is taken to be simply desvribed by the quaratic dispertion without spin orbtit effects. the simple model with quadratic dispersion without spin orbtit effects.
The electron eigenstate wavefunction $\Psi$ with the energy $E$ satisfies the following Schr\"odinger equation:
\begin{equation}\label{eq_Ham}
\Bigl( \frac{{\bf p}^2}{2 m_{\ast}} - \alpha {\bf M}({\bf r}) \cdot \hat{\bf S} \Bigr) \Psi = E\Psi,
%H = \frac{{\bf p}^2}{2 m} - g {\bf M}({\bf r}) \cdot \hat{\bf S}
\end{equation}
where ${\bf p}$ is the 2D momentum operator, $\hat{\bf S}$ is the electron spin operator, $m_{\ast}$ is the electron in-plane effective mass and $\alpha$ is the exchange coupling constant.

Let us analyze the asymptotic of $\Psi$. Outside of the vortex core the itinerant carrier is embedded into the homogeneous magnetization environment ${\bf M} = \eta M {\bf e}_z$, which gives rise to the carrier band spin splitting $\Delta = \alpha M$ ($\eta = \pm 1$ is the background magnetization direction normal to the
film plane outside of the core).
We assume $\Delta/2E<1$ so that both spin subbands are activated (the electron energy $E$ is of the order of Fermi energy).
We introduce the adiabatic parameter $\lambda_a$ in the form:
\begin{equation}
\label{eqLambda}
\lambda_a = a k (\Delta/2E),
\end{equation}
where $k=\sqrt{2 m_{\ast} E/\hbar^2}$ and $a$ is the vortex size.
Due to background magnetization outside of the vortex core the spin-down and spin-up states with the same energy have different wavevectors: \begin{equation}\label{eqdiffk} k_{\uparrow, \downarrow}^2 = 2m_{\ast}(E \pm \eta \Delta/2) / \hbar^2 .\end{equation}
%$E$ has the meaning of Fermi energy. here we interested in case, when two spin subbands are under the  $E>\Delta$.
%we define the adiabatic parameter $\lambda_a = (a k) \Delta/E$, where $k=\sqrt{2 m E/\hbar^2}$, $a$ is the skyrmion radius.
%In the asymptotic region ($k r \gg 1$) the wavevector for the given energy $E$ differs for spin up and down states: $k_{\uparrow, \downarrow}^2 = 2m(E \pm \eta \Delta) / \hbar^2  $.
%Taking this into account the asymptotic form of wave function $\Psi$ for the scattering problem is given by:
%If the incident plane wave goes along $x$ axis,
Far from the vortex core ($kr\gg1$) the wavefunction is given by: %they have the asymptotic forms:
% Let the incident plane wave goes along $x$ axis. With this taken into account, the wave function $\Psi$ has the asymptotical form (large distances $kr\gg1$):
\begin{equation}\label{eq_PSI}
\Psi =
\begin{pmatrix}
e^{i k_{\uparrow} x } u_{\uparrow}
\\
e^{i k_{\downarrow} x } u_{\downarrow}
\end{pmatrix}
+
\frac{1}{\sqrt{r}}
\begin{pmatrix}
e^{i k_{\uparrow} r } \bigl( f_{\uparrow \uparrow} u_{\uparrow} + f_{\uparrow \downarrow} u_{\downarrow} \bigr)
\\
e^{i k_{\downarrow} r }  \bigl( f_{\downarrow \uparrow} u_{\uparrow} + f_{\downarrow \downarrow} u_{\downarrow} \bigr)
\end{pmatrix},
\end{equation}
where the first term is the incident plane wave and the second term is the outgoing cylindrical wave, $u = (u_1, u_2)^{T}$ is the incoming wave polarization spinor ($|u_1|^2 + |u_2|^2 =1$), $f_{\alpha \beta}(\theta)$ is the scattering amplitude, $\theta$ is the scattering angle, it is also the polar angle in the coordinate system used as the incident plane wave is assumed coming along the $x$-axis. There are four scattering channels: two spin-conserving channels $\left|\uparrow\right \rangle\rightarrow\left|\uparrow\right\rangle$, $\left|\downarrow\right \rangle\rightarrow\left|\downarrow\right\rangle$, and two spin-flip channels
$\left|\uparrow\right \rangle\rightarrow\left|\downarrow\right\rangle$, $\left|\downarrow\right \rangle\rightarrow\left|\uparrow\right\rangle$.
The partial differential scattering cross sections for each channel are given by \begin{equation}
\label{eqSigmaDiff}
\frac{d\sigma_{\alpha \beta}}{d\theta }= \frac{k_{\alpha}}{k_{\beta}} |f_{\alpha \beta}(\theta)|^2.
\end{equation}

%\textcolor{red}{UNCLEAR THAT WE ARE CONSIDERING 2D CASE}
We proceed with discussing of the magnetization field ${\bf M}({\bf r}) = M {\bf n}({\bf r})$, where ${\bf r}=(x,y)$ is an in-plane radius vector,
${\bf n}$ is a unit vector describing the spatial dependence of the magnetization direction.
We introduce the commonly used parametrization for the chiral magnetization field
 ${\bf n} = (\sin{\Lambda} \cos{\Phi}, \sin{\Lambda} \sin{\Phi}, \eta \cos{\Lambda})$, where the vortex profile $\Lambda(r)$ depends on the in-plane radius vector magnitude $r$, $\Phi(\theta) = \kappa \theta + \gamma$, where $\theta$ is the polar angle.
 The nonzero value of the spin chirality is described by the integer parameter $\kappa$ called vorticity, which determines the direction of the in-plane twist,
  helicity $\gamma$ determines the initial phase of this rotation.
% the integer parameter $\kappa$ called vorticity indicates determines the direction of the in-plane twist.
 The vortex perpendicular orientation $\eta = \pm 1$ denotes the background magnetization direction normal to the
 film plane outside of its core (we assume that the sign of $\cos{\Lambda(r \rightarrow \infty)} = +1$ is fixed).
  %is the   \textcolor{red}{MAYBE COMMENT ON ITS SENSE}.
The topological characteristic of such a structure is the topological charge (also known as winding number):
 %Various configurations are topologically distinguished by possessing a different topological charge:
%Topologically nonequivalent configurations possess
\begin{equation*}
Q = \frac{1}{4 \pi} \int {\bf n} \left(\partial_x {\bf n} \times \partial_y {\bf n} \right)d{\bf r} = \eta \frac{\kappa }{2}\left( \left. \cos{\Lambda} \right|_{\infty} - \left. \cos{\Lambda} \right|_0 \right).
\end{equation*}
A topologically nontrivial structure of the magnetization field $Q \neq 0$ is called magnetic skyrmion.
It has an
%; its hallmark is the
opposite orientation of magnetization inside and outside of its core.
A topological Hall response in a system with magnetic skyrmions has been considered in a meanfield approximation~\cite{Ye1999, Onoda_SkyrmNumber}.
%as a basic source of the
%topological Hall response via
%in a number of systems, where the carriers are influenced by the averaged behaviour of magnetization\cite{BrunoDugaev}. %carriers must
%ordinary meanfield description of topological fields reffers to skyrmion as a basic source of Hall response.
%for Hall response
%The skyrmions are essential to produce the nonzero topological field within the meanfield description.
%In what follows we will for the concretenesse consider the case of symmetric magnetic skyrmion ($Q\neq 0$, $\kappa=+1$).
%Here we solve the scattering problem exactly,
On the contrary, the method we use in our work is exact and accounts for the local character of interaction during the scattering. The emergence of Hall response is due to local noncollinear ordering of magnetic moments. Therefore, even chiral configurations with zero winding number (e.g. co-vortices \cite{Fraerman1} which have $Q=0$, but $\kappa\neq0$) should also produce a transverse scattering contributing to THE.

%To be specific, below we mainly focus on the case of an axially symmetric magnetic skyrmion ($Q\neq 0$, $\kappa=+1$), while the case of $Q=0$ will be discussed qualitatively.
%Our goal is to calculate the scattering amplitude $f_{\alpha \beta}(\theta)$.
%\textcolor{blue}{
In order to calculate the scattering amplitude $f_{\alpha \beta}(\theta)$ let us introduce a set of basis states for the considered scattering problem.
We notice that the specific angular dependence of chiral magnetic vortex allows one to separate $r$ and $\theta$ in Eq.(\ref{eq_Ham})~\footnote{The corresponding operator commuting with a total Hamiltonian in Eq.(\ref{eq_Ham}) is given by $- i \partial_{\theta} + \kappa \hat{S}_z$}.
%, which simplifies the finding of $f_{\alpha \beta}(\theta)$.
Indeed, the explicit form of the scattering potential $V_{sc}$ due to an electron exchange interaction with the
%introduced above classical field of
magnetization field ${\bf M}({\bf r})$ is given by:
%\begin{equation}
%\label{eq_Pot}
%V_{sk} = -\frac{\Delta}{2}
%\begin{pmatrix}
%- \eta (1-\cos{\Lambda(r)}) & e^{- i \kappa \theta - i \gamma} \sin{\Lambda(r)}
%\\
%e^{ i \kappa \theta + i \gamma} \sin{\Lambda(r)} & \eta (1-\cos{\Lambda(r)})
%\end{pmatrix}.
%\end{equation}
\begin{equation}
\label{eq_Pot}
V_{sc} = -\frac{\Delta}{2}
\begin{pmatrix}
- \eta (1-n_z(r)) & e^{- i \kappa \theta - i \gamma} n_{\parallel}(r)
\\
e^{ i \kappa \theta + i \gamma} n_{\parallel}(r) & \eta (1-n_z(r))
\end{pmatrix},
\end{equation}
where $n_{z}(r) = \cos{\Lambda(r)}$, $n_{\parallel}(r) = \sin{\Lambda(r)}$.
The off-diagonal components of $V_{sc}$ mixing spin-up and spin-down states contain an additional angular factor $e^{i \kappa \theta}.$
% ($\kappa$ is integer).
Hence, the Hamiltonian (\ref{eq_Ham}) eigenstates can be labeled by an angular momentum projections $m$ with the angular part of the eigenfunction given by a combination of $e^{i m \theta} |\uparrow \rangle$ and $e^{i (m + \kappa) \theta + i \gamma} |\downarrow \rangle$ states (see details in Appendix A).
%}
%(is connected with a fact, that the operator $- i \partial_{\theta} + \kappa \hat{S}_z$ commutes with total Hamiltonian).
Taking this into account $f_{\alpha \beta}(\theta)$ is written in the form:
\begin{widetext}
	\begin{equation}\label{eq_f}
	%{f}_{\alpha \beta} = \frac{1}{i \sqrt{2 \pi k_{\alpha}}}
	%\sum_m e^{i m \varphi} d_{\alpha} \left(\hat{S}_m - \hat{I}\right)_{\alpha \beta}
	{f}_{\alpha \beta}(\theta) = \frac{1}{\sqrt{2 \pi i k_{\uparrow}}  } \sum_m e^{i m \theta}
	\begin{pmatrix}
	S_m^{\uparrow \uparrow} - 1 &   S_m^{\uparrow \downarrow}
	\\
	e^{i \kappa \theta  + i \gamma} \sqrt{\frac{k_{\uparrow}  }{k_{\downarrow} }} S_m^{\downarrow \uparrow} & e^{i \kappa \theta + i\gamma} \sqrt{\frac{k_{\uparrow}  }{k_{\downarrow} }} \left(S_m^{\downarrow \downarrow} -1 \right)
	\end{pmatrix}_{\alpha \beta},
	\end{equation}
\end{widetext}
where ${S}^{\alpha\beta}_m$ are the partial scattering matrices.
%determined from the asymptotic behavior of ${g}_m^{\nu}$.
${S}^{\alpha\beta}_m$ are computed for an arbitrary adiabatic parameter $\lambda_a$ using the phase-functions method (the details are given in Appendix A).
%(the details of computation are given App.A).
%The formula (\ref{eq_f}) gives the scattering amplitude for an arbitrary adiabatic parameter $\lambda_a$.
The dependence of the differential cross section (\ref{eqSigmaDiff})
%$d\sigma_{\alpha \beta}/d\theta = \left(k_{\alpha}/k_{\beta}\right) |f_{\alpha \beta}(\theta)|^2$
on $\lambda_a$ is the main subject of the following section.
%Its dependence on $\lambda_a$ is the main subject of the following section.

\section{Analysis of scattering}
%We proceed to discuss the scattering throughout the full range of $\lambda_a$.
%In presented below numerical results

\subsection{General properties of asymmetric scattering}
Let us mention some general aspects of the scattering on a magnetic skyrmion.
%In general each of the scattering channels possesses a transverse response, and all processes contribute to the net Hall %effect. Indeed,
As equation (\ref{eq_Pot}) suggests the harmonics with the opposite angular momentum projections $m$ and $-m$ are not identical, so that ${S}_m \neq {S}_{-m}$, and the cross section $d \sigma_{\alpha \beta}$ gets an asymmetric contribution.
Let us divide $d \sigma_{\alpha \beta}$ into symmetric and antisymmetric parts:
\begin{equation}\label{eq_dsigma}
\frac{d \sigma_{\alpha \beta}}{d \theta} = \frac{k_{\alpha}}{k_{\beta}} |f_{\alpha \beta}|^2 = G_{\alpha \beta}(\theta) + \Sigma_{\alpha \beta}(\theta),
\end{equation}
where $G_{\alpha \beta}(\theta) = G_{\alpha \beta}(-\theta)$ is symmetric, and $\Sigma_{\alpha \beta}(\theta) = - \Sigma_{\alpha \beta}(-\theta)$ is antisymmetric with respect to the scattering angle $\theta$. It is the antisymmetric part $\Sigma_{\alpha \beta}$ that gives rise to the net perpendicular current and the following Hall response.

%In this section we discuss
The properties of the asymmetric scattering $\Sigma_{\alpha \beta}$ strongly depend on the adiabatic parameter $\lambda_a$.
% (this dependence is a subject of the following subsections).
%It is exactly the terms $\hat{\Sigma}$, which describes the describing phenomena of .
%The antisymmetric part $\Sigma_{\alpha \beta}$ leads to the net perpendicular current,it appears that its properties strongly depend on the adiabatic parameter $\lambda_a$.
The latter is expressed as a product of two dimensionless parameters $ka$ and $\Delta/2E$ (\ref{eqLambda}).
%\textcolor{blue}{
%The $\Sigma_{\alpha \beta}$ produces the Hall response, and its dependence on $\lambda_a$ is the main subject of this %work.}
%Thus, with the electron energy $E$ being fixed, the adiabatic parameter can be varied either by changing the exchange splitting $\Delta$ or by varying the skyrmion size $a$.
%While it is the magnitude of $\lambda_a$ that controls the applicability of adiabatic or perturbative approach to the scattering on a skyrmion, it is important which of the two parameters is varied.
The adiabatic parameter discriminates between two qualitatively different regimes leading to either spin Hall effect or charge Hall effect in the electron scattering on a skyrmion.
%The magnitude of the adiabatic parameter determines the validity of the Born approximation, which turns to be the key origin of the crossover between different spin asymmetry of the scattering cross-section. Before discussing the scattering asymmetry crossover we note that beyond the adiabatic parameter some details of the scattering are independently determined by the the parameters contributing to $\lambda_a$.
The magnitude of $k a$
determines the number of angular harmonics contributing to the scattering;
$ka/2\ll1$ corresponds to s-scattering\cite{LL3},
%(the cross-section is approximately the same for all scattering angles)
while in the case
%
%amplitude $f_{\alpha \beta}$ in (\ref{eq_f}). For small $ka$ the very first harmonics are activated and we have the %s-type scattering; if
$ka/2 > 1$ multiple angular harmonics contribute to $f_{\alpha \beta}$ so that the scattering cross section (\ref{eq_dsigma}) has a complex dependence on the scattering angle $\theta$ mostly in the range of small $\theta$. Also, the larger is the skyrmion size (and $ka$) the larger is the magnitude of the cross section.
%and it is highly localized near the region
%the scattering angle .
Regarding the role of $\Delta/2E$,
%in the magnitude of $\lambda_a$
when the exchange splitting exceeds the scattering energy $\Delta/2E >1$, only one spin subband contributes to the transport and the initial electron state becomes fully polarized. Obviously, in this limiting case the transverse spin current and charge current are equal regardless the skyrmion size $a$ and the value of $\lambda_a$.
We will not discuss this regime.

Let us clarify the role of the vortex parameters $\kappa$, $\gamma$ and $\eta$ on the asymmetric part of the scattering cross section  $\Sigma_{\alpha \beta}$.
The helicity $\gamma$ enters $f_{\alpha \beta}$ only through a common phase factor $e^{ i \gamma}$, hence it doesn't affect the scattering cross section $d\sigma_{\alpha \beta}/d\theta \sim |f_{\alpha \beta}|^2$; in what follows we take $\gamma = 0$ for simplicity.
The sign of the vorticity $\kappa$ determines the sign of the scattering asymmetry. Indeed, the replacement
 $\kappa \rightarrow - \kappa$ leads to $\Sigma_{\alpha \beta}(\theta) \rightarrow \Sigma_{\alpha \beta}(-\theta) = -\Sigma_{\alpha \beta}(\theta)$.
 %; hence the sign of the vorticity $\kappa$ determines the resulting sign of $\Sigma_{\alpha \beta}(\theta)$.
As will be further discussed below the role of the background magnetization $\eta$ appears to be not so trivial. There is a global property related to $\eta$: $\Sigma_{\alpha \beta}(\theta, \eta) = \Sigma_{\bar{\alpha} \bar{\beta}}(-\theta,-\eta)$, where $\bar{\alpha}$ denotes the spin state opposite to $\alpha$. However, whether the type of asymmetry (sign of $\Sigma_{\alpha \beta}(\theta)$) changes upon $\eta \rightarrow - \eta$ depends on the scattering regime (small or large $\lambda_a$). %\textcolor{red}{
%The physics of the scattering in these
%We'll address this in depth within the consideration of the corresponding regime.
%}

\begin{figure*}
	\includegraphics[width=1.\textwidth]{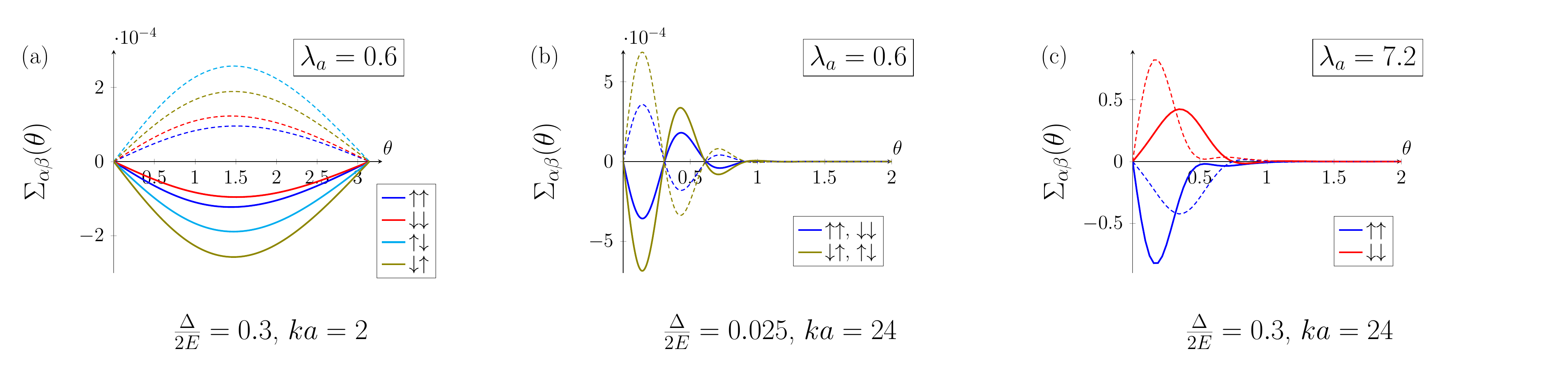}%{pict_4-2.pdf}%{fig2.pdf}%{pict_4_hor.pdf}%{fig-cr-2.eps}%{fig-1.pdf}%{fig-1.png}%{3-3.jpg}%{BDS.eps}
	\caption{
		Asymmetric contribution to the electron cross section $\Sigma_{\alpha \beta}(\theta)$ (in units of skyrmion radius $a/2$) on a single magnetic skyrmion as a function of scattering angle $\theta$ for small (a,b) and large (c) values of the adiabatic parameter. Solid and dashed lines correspond to the skyrmion orientation $\eta = 1$ and $\eta = -1$, respectively.
	}	
	\label{figan1}
\end{figure*}

We would like to highlight that the particular shape of a magnetic vortex profile $n_{z}({ r})= \cos{\Lambda(r)}$
%Regarding the role of the vortex profile $n_{z}(r) = \cos{\Lambda(r)}$ we state that the shape of $n_{z}(r)$
has rather quantitative effect on the scattering. Moreover, whether $\Lambda(r)$ describes a topologicaly nontrivial magnetic skyrmion ($\Lambda_{0} \neq \Lambda |_{\infty}$ and $Q \neq 0$) or trivial magnetic vortex ($\Lambda_{0} = \Lambda |_{\infty}$ and $Q = 0$) doesn't have any qualitative concequences on $\Sigma_{\alpha \beta}(\theta)$ behaviour.
%\textcolor{red}{
In particular,
%Particularly,
%\textcolor{red}{
the cross section asymmetry is due to
%related to
a nonzero integer $\kappa$ rather than
%and not
to the concrete form of $n_z(r)$.
To be specific, we firstly focus on
%provide the detailed analysis of
%scattering for
a magnetic skyrmion with $\kappa=+1$.
Scattering on a topologicaly trivial strucuture $Q=0$
%\textcolor{red}{
is considered in section \ref{secTriv}.
We demonstrate that it has properties similar to those of the scattering on topologically non-trivial skyrmions.
%It exhibits a similar behavior to that
%will be considered further.
%}
\subsection{Scattering on a magnetic skyrmion with $\kappa = +1$}
In this section we consider a skyrmion with $\kappa = +1$ of a finite radius, so that for $r>a/2$ there is no perturbation of magnetization over its uniform background value. Inside its core $r<a/2$ the skyrmion was parameterized via $\Lambda(r) = \pi \sin^2 \left(\pi/2(1+2r/a)\right)$.
%We found no dependence of THE on skyrmion helicity, so for simplicity we take $\gamma = 0$.

Fig.~\ref{figan1} shows the asymmetric part of the scattering cross section $\Sigma_{\alpha \beta}(\theta)$ computed using equations (\ref{eq_f}) and (\ref{eq_Ap1}).
% for the limiting cases.
Positive (negative)  values of $\Sigma_{\alpha \beta}(\theta)$ for the positive scattering angle $\theta$ in Fig.~\ref{figan1} correspond to the preferable scattering to the left (to the right) with respect to the incident flux direction. Figs.~\ref{figan1}a,b illustrate the scattering for a small $\lambda_a$ while the case of a large $\lambda_a$ is shown in Fig.~\ref{figan1}c, in both cases $\Delta/2E < 1$. The results are shown for the two opposite skyrmion orientations $\eta=\pm 1$ (solid and dashed lines respectively).

%\subsubsection{Scattering in the limiting regimes}
\subsubsection{Weak coupling regime}
%\subsection{Scattering in the limiting cases}
The weak coupling regime corresponds to a small magnitude of the adiabatic parameter $\lambda_a<1$.
%We first analyze the scattering for small $\lambda_a$. The
 Fig.~\ref{figan1}a illustrates the case of s-type scattering ($k a\sim 1$). It is clearly seen that each scattering channel has the same sign of $\Sigma_{\alpha \beta}(\theta)$, so both spin-up and spin-down electrons
 %disregard to its initial and final spin state is
 are preferably scattered into the same half-plane.
 In this regime the transverse charge current clearly dominates over the spin current and the topological Hall effect leads to a pronounced transverse charge current even for non-polarized electrons.
 %of a charge Hall response doesn't require nonzero spin polarization.

\begin{figure}
	\centering
	\includegraphics[width=0.35\textwidth]{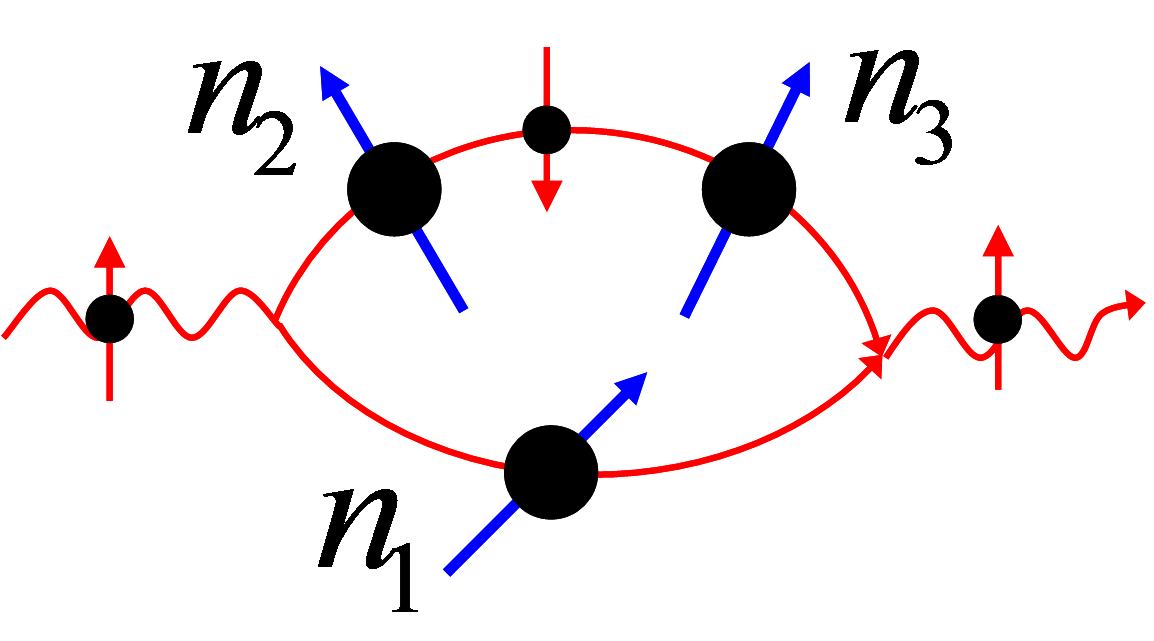}%{fig3.pdf}
	\caption{An electron scattering on a three non-coplanar spins. The diagram is given for spin-conserving (spin-up) scattering channel and shows the interference between spin-conserving scattering on scatterer 1 with magnetization direction ${\bf n_1}$ and double spin-flip process on scatterers 2,3 with ${\bf n}_{2,3}$. The asymmetry of arises from nonzero spin chirality $ {\bf n}_1 \cdot \left[ {\bf n}_2\times {\bf n}_3\right] \neq 0$.}
%Interference betw}
	\label{figtriada}
\end{figure}

From the symmetry point of view this effect is similar to the ordinary Hall effect;
%This can be explained similarly to the ordinary Hall effect: from the symmetry consideration
the presence of $z$-component of a pseudovector breaking the time reversal symmetry leads to a transverse pure charge current when an electric current flows along the sample.
%\textcolor{red}{
Non-collinear magnetic textures have a non-zero spin chirality that is a combination of three non-collinear spins forming the magnetization field $\chi_{123} = {\bf n}_1 \cdot \bigl[{\bf n}_2 \times {\bf n}_3\bigr]$,
where the spatial positions of the sites $1,2,3$ are arranged in the clockwise direction (Fig.\ref{figtriada}) and ${\bf n}_i$ is the local direction of magnetization at i-th site.
%with fixed direction of the circuit in space (e.g. clockwise)
%\textcolor{red}{(with fixed direction of the circuit in space, e.g. clockwise)}
The chirality breaks time reversal symmetry and behaves as $z$-component of a pseudovector under mirror-reflections; so does any linear combination of $\chi_{ijk}$ for different space points $i,j,k$.
%combination of 3 non-collinear spins $ {\bf M}_1 \cdot \bigl[{\bf M}_2 \times {\bf M}_3\bigr]$ with fixed direction of the circuit in space (e.g. clockwise), that is called spin chirality $\chi_{123}$,
At a small $\lambda_a$ when the Born series expansion is applicable, the dominant term contributing to $\Sigma_{\alpha \beta}$ is directly related to a linear combination of $\chi_{ijk}$\cite{prl_skyrmion}, thus in the weak coupling regime the magnetization field chirality is analogous to a magnetic field acting on spinless particles and producing a transverse charge current.
The chiral symmetry allows for the existence of such an effective magnetic field for arbitrary
$\lambda_a$, in consistence with our finding that the transverse charge current persists up to
$\lambda_a\sim 1$.

%As the chirality induced effective magnetic field
%\textcolor{red}{
%We would like to notice that this hallmark is
%not only valid in the limit $\lambda_a \rightarrow 0$ but persists as well for higher $\lambda_a \le 0.8 $ (see %Figs.~\ref{figdisc1},\ref{figHall1}); the domination of charge current is protected by rigorus symmetry arguments. 	
%}
%According to symmetry arguments this generates a transverse charge current; an electron is scattered to the same half-plane disregard to its spin.
%as is observed in Fig.~\ref{figan1}.
%}appears to be the same

In order to shed some light on the appearance of the same scattering asymmetry for spin-up and spin-down electrons
let us consider a scattering of an electron on a triad of non-coplanar spins (Fig.~\ref{figtriada}).
The details of the calculations are presented in Appendix \ref{AppTriad}.
%Let us first consider the spin conserving scattering channels.
For spin conserving scattering channels
%A nonzero
the spin chirality
%${\bf M}_1 \cdot \bigl[{\bf M}_2 \times {\bf M}_3\bigr]$
manifests itself in the interference between
%single
spin-conserving scattering on one of the magnetic centers in the triad and
%the first
%scattering via
double spin-flip scattering on the other two (Fig.~\ref{figtriada}).
The key feature of this interference is that its contribution
  to the asymmetric part of the cross section has the same sign for spin-up and spin-down diagonal scattering channels  ($\Sigma _{ \downarrow  \downarrow }$ and $\Sigma _{ \uparrow  \uparrow }$ ).
  % have the same sign.
This is because the
%occurs from the fact that the
%effect
%arises from the mutual compensation of
%signs between one spin-conserving and two sin-flip processes.
opposite signs in the matrix elements for spin-conserving scattering of spin-up and spin-down electron on  scatterer 1 %$V_{1\uparrow\uparrow}=-V_{1\downarrow\downarrow}$ (see Eq.~\ref{eq_Pot} )
are compensated by the sign change for the double spin-flip scattering on scatterers 2,3
%:  ${\mathop{\rm Im}\nolimits} \left({V_2}_{ \uparrow  \downarrow }{V_3}_{ \downarrow  \uparrow }\right) =  - %{\mathop{\rm Im}\nolimits} \left({V_2}_{ \downarrow  \uparrow }{V_3}_{ \uparrow  \downarrow }\right)$
(see Appendix \ref{AppTriad}).
Exactly the same effect appears for non-diagonal (spin-flip) scattering channels $\Sigma _{ \uparrow  \downarrow },\Sigma _{ \downarrow\uparrow   }$.
  \textcolor{red}{
%a mutual compensation of opposite signs within each channel: according to Eq.~\ref{eq_Pot} spin conserving single scattering event is described by similar matrix elements having opposite signs: $V_{11} = - V_{22}$, second order spin flip processes are takein in different order producing additional changing of overall sign.
%is the interference between first order spin conserving and second order double spin-flip scattering (Fig.~\ref{figtriada}), this interference produces the same sign for the asymmetrical chirality-aware part of the cross section both for spin up and spin down electrons.
}

While the asymmetry in the weak coupling regime is the same for spin-up and spin-down electrons, the asymmetrical cross section also depends on the skyrmion size.
%Let us clarify the role of skyrmion size in the weak coupling regime.
For small $ka$ the cross section is determined by the lowest angular harmonics so the asymmetric part takes the form $\Sigma_{\alpha \beta}(\theta) \sim \sin{\theta}$ (Fig.~\ref{figan1}a). When multiple angular harmonics are involved ($ k a \gg 1$), the asymmetric part of the cross section oscillates with the scattering angle as shown in Fig.~\ref{figan1}b. Increasing the skyrmion size while keeping $\lambda_a \ll 1$ suppresses both spin and charge transverse currents due to oscillating structure of
%emergent
$\Sigma_{\alpha \beta}(\theta)$. In Fig~\ref{figan1}b the relations $\Sigma_{\uparrow \uparrow} = \Sigma_{\downarrow \downarrow}$, $\Sigma_{\uparrow \downarrow} = \Sigma_{\downarrow \uparrow}$ hold in agreement with the perturbation theory~\cite{prl_skyrmion} in the limit of $\Delta/E \rightarrow 0$.

%In both considered cases of a small $\lambda_a$
%the scattering asymmetry is the same for spin-flip scattering channels and spin-conserving channels (according to the perturbation theory \cite{prl_skyrmion} the following relation $\Sigma_{\uparrow \uparrow} = \Sigma_{\downarrow \downarrow}$, $\Sigma_{\uparrow \downarrow} = \Sigma_{\downarrow \uparrow}$ hold in the limit of $\Delta/E \rightarrow 0$).
In the weak coupling regime the contribution to the asymmetric scattering from spin-flip processes always prevails over that from spin-conserving ones.
%For the weak coupling regime
Reversing the background magnetization (skyrmion orientation) sign $\eta \rightarrow - \eta$ changes the preferred transverse scattering direction for each scattering channel and therefore changes the sign of the Hall effect. In Figs.~\ref{figan1}a,b the asymmetric part $\Sigma_{\alpha \beta}(\theta)$ for $\eta = - 1$ is plotted in dashed lines.
%The sign of the resulting Hall effect is determined by a skyrmion orientation in this %regime.
%As can be also seen from Figs.~\ref{figan1}a,b reversing skyrmion orientation $\eta$ changes the preferred transverse scattering direction for each scattering channel.

\subsubsection{Adiabatic regime}

Fig.~\ref{figan1}c corresponds to the adiabatic regime $\lambda_a \gg 1$. Here only spin-conserving terms are shown since spin-flip scattering is suppressed.%in this case
As can be clearly seen in Fig.~\ref{figan1}c, spin-up and spin-down electrons have different scattering asymmetry, they are preferably scattered into the opposite half-planes creating a transverse spin current. This feature is described by the adiabatic Berry phase theory, which allows to reduce the scattering on a skyrmion to the action of an effective magnetic field having opposite sign for spin-up and spin-down electrons. According to this mechanism a finite spin polarization of the carriers is necessary to convert spin Hall effect into a nonzero transverse charge current\cite{BrunoDugaev,Spin_Top_Hall}.

Unlike the case of a small $\lambda_a$, the type of the scattering asymmetry in Fig.~\ref{figan1}c is determined solely by the electron initial spin state (for a fixed vorticity $\kappa$),
%the inversion of the skyrmion orientation $\eta \rightarrow - \eta$ does not change the type of scattering asymmetry for %a given scattering channel,
i.e. spin-up electrons scatter to the left regardless skyrmion orientation $\eta=\pm 1$. This behavior is also in agreement with the explanation given by the adiabatic theory. The effective magnetic field associated with the geometric Berry phase acquired by the electron wavefunction moving through the spin texture is opposite for spin-up and spin-down states as they are at the opposite poles of the Bloch sphere\cite{BrunoDugaev}. The background magnetization inversion $\eta \rightarrow - \eta$ doesn't swap the electron spin-up and spin-down states on the Bloch sphere and hence the sign of the effective magnetic field is not changed.
%It should be noted that
%while the effective magnetic field is the same for spin up and spin down states,  the actual cross sections for spin-up and spin-down electrons differ due to their dependence on the kinetic energy of the incident particle ($k_{\uparrow} \neq k_{\downarrow}$).

Since $\Delta/2E <1$ the adiabatic regime $\lambda_a \gg 1$ is achieved only when $k a\gg 1$. Therefore, the spin Hall effect for $\lambda_a \gg 1$ with both spin subbands activated always involves many angular harmonics in the scattering.
At $\lambda_a\gg1$ the Born approximation is invalid, at that the quasi-classical approach becomes more adequate when an electron is treated as a localized wavepacket adiabatically moving in a smooth magnetization field.
%consisiting of a bunch of different $k$.
%\textcolor{red}{
%While in the weak coupling regime described by quantum Born series, we see that at large $\lambda_a$ the alternative %quasiclassical approach becomes more adequate when an electron is treated as a localized wavepacket consisiting of a %bunch of different $k$.
%Then the asymmetrical scattering arises from interference between different trajectories of a localized wavepacket moving in a chiral magnetization field. %In this regime the scattering asymmetry appears to be different for spin up and spin down electrons.
%}
%wave-pocket, quasiclassical picture	

\subsubsection{Crossover}

\begin{figure*}
	\centering
	\includegraphics[width=1.\textwidth]{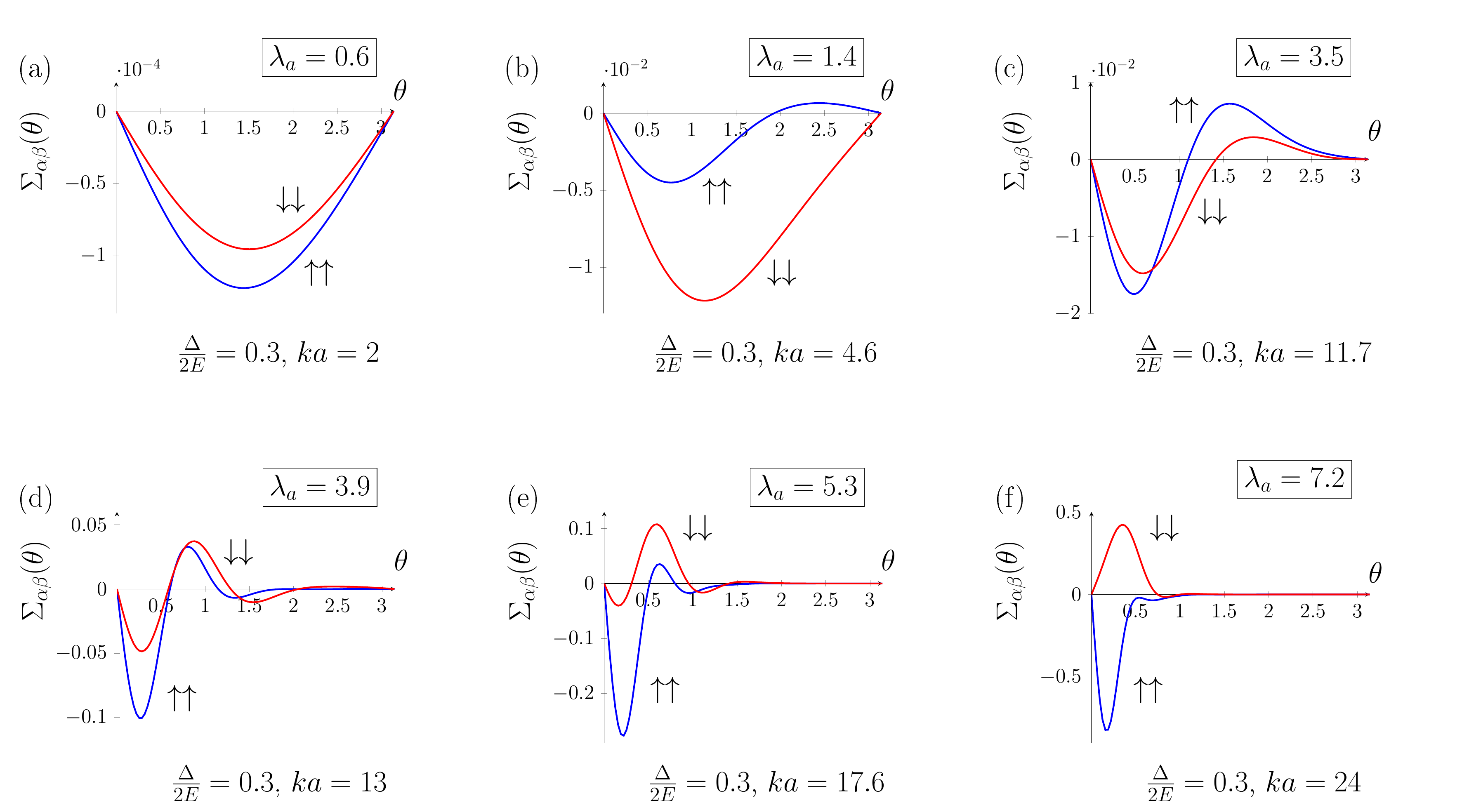}%{fig3.pdf}
	\caption{Evolution of $\Sigma_{\uparrow \uparrow},\Sigma_{\downarrow \downarrow}$ (in units of $a/2$) tuned by skyrmion size $a$. }
	\label{figcr1}
\end{figure*}

\begin{figure*}
	\centering
	\includegraphics[width=1.\textwidth]{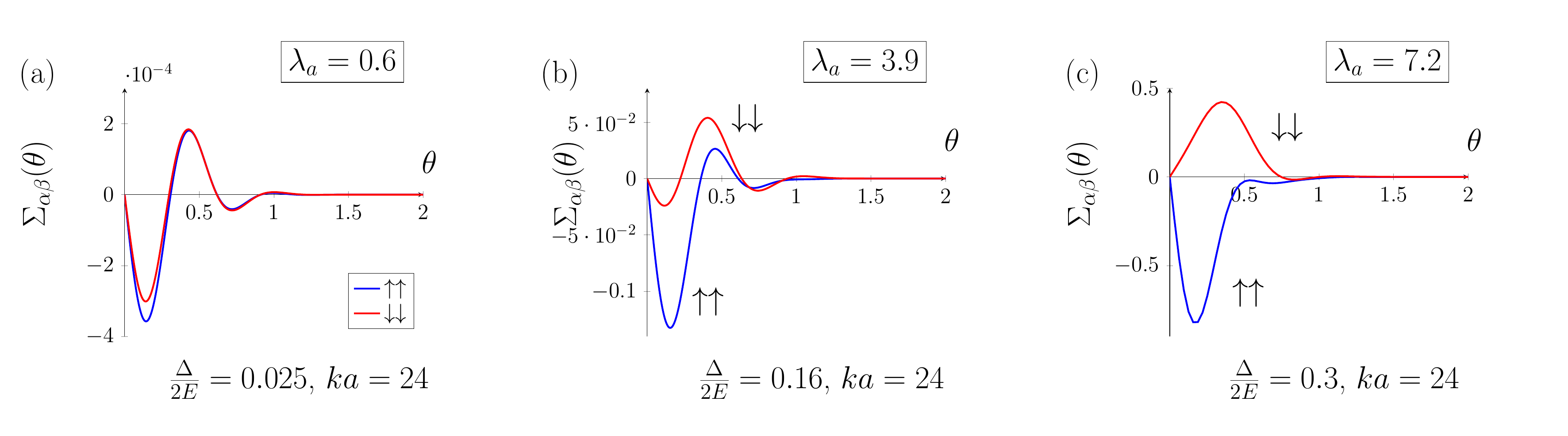}%{fig3.pdf}
	\caption{Evolution of $\Sigma_{\uparrow \uparrow},\Sigma_{\downarrow \downarrow}$ (in units of $a/2$) tuned by an exhange coupling $\Delta$. }
	\label{fig3b}
\end{figure*}

%Note, that this effect holds only while the second Born approximation is valid. The increase of the adiabatic parameter suppresses the spin flip processes as spin up and spin down electrons of equal energy now have significantly different wavevectors and Green's functions describing their free propagation.

%The deviation from the weak coupling regime also makes the scattering picture completely different. For a fixed wavevector $k$ one should account for higher order Born series, at that the alternative quasiclassical approach becomes more adequate when an electron is treated as a localized wavepacket consisiting of  a bunch of different $k$. Then the asymmetrical scattering arises from interference between different trajectories of a localized wavepacket moving in a chiral magnetization field. In this regime the scattering asymmetry appears to be different for spin up and spin down electrons.

Let us now discuss evolution of asymmetric scattering between the weak coupling and adiabatic regimes.
%The increase of the adiabatic parameter has two consequences.
%Firstly,
As $\lambda_a$ is getting larger the spin-flip processes get suppressed as spin-up and spin-down electrons of equal energy now have significantly different wavevectors (\ref{eqdiffk}), this results in rapidly oscillating factors in the spin-flip scattering matrix elements. Consequently, the spin-independent contribution to the asymmetric scattering vanishes.
Moreover, the scattering picture becomes completely different:
%the deviation from the weak coupling regime makes
%higher order Born series,
%For a fixed wavevector $k$
moving away from the weak coupling regime one should account for higher order Born series. Then the scattering is better described treating an electron as
 %the alternative quasiclassical approach becomes more adequate when an electron
 %is treated
 a spatially localized wavepacket
moving in the effective magnetic field due to Berry curvature which has different sign for spin-up and spin-down electron states.
%quasiclassical wavepacket
%consisiting of  a bunch of different $k$.
%Then the asymmetrical scattering arises from interference between different trajectories of a localized wavepacket moving %in a chiral magnetization field. In this regime the scattering asymmetry appears to be different for spin up and spin %down electrons.
%and functions describing their free propagation.
%\textcolor{red}{
%The crossover
%between different types of scattering asymmetry
%is driven by the transition from the interference of the first two Born terms characterized by the same type of %scattering asymmetry to a quasi-classical picture of scattering with suppressed spin flip transitions and opposite type %of scattering asymmetry.
%Increasing the adiabatic parameter one should account for higher order Born series which triggers a formation of %quasiclassical wavepacket moving in the effective Berry's phase magnetic field.
%at that the alternative quasiclassical approach becomes more adequate when an electron is treated as a localized wavepacket consisiting of a bunch of different $k$.
%Then the asymmetrical scattering arises from interference between different trajectories of a localized wavepacket moving in a chiral magnetization field. In this regime the scattering asymmetry appears to be different for spin up and spin down electrons.
%}

%As it was mentioned above, the crossover between different types of scattering asymmetry is driven by the transition from the Born approximation to a quasi-classical picture of scattering with suppressed spin flip scattering.

The details of the crossover appear to be different depending on whether the exchange interaction strength $\Delta$ is varied keeping  $ka$ constant or the skyrmion size $a$ is varied keeping the exchange strength fixed (\ref{eqLambda}).
The evolution of the asymmetrical part of the differential cross section $\Sigma_{\alpha \beta}(\theta)$
%as a function of the scattering angle $\theta$ tuned
with the skyrmion size is shown in Fig.~\ref{figcr1}.
%The asymmetrical part of the differential cross-section $\Sigma_{\alpha \beta}(\theta)$ as a function of the scattering angle $\theta$ is shown for different $\lambda_a$.
%is shown in Fig.~\ref{figcr1}
%(only spin flip channels are shown for the purpose of clarity).
Only spin conserving channels are shown for the purpose of clarity.
%within the sequence in
The first and the last frames in Fig.~\ref{figcr1}a,f correspond to the limiting cases considered in the previous section (see Fig.~\ref{figan1}a,c).

%We first duscuss the transition that the asymmetric part $\Sigma_{\alpha \beta}(\theta)$ undergoes with changing $\lambda_a$.
%There are some features of the transition between limiting regimes that we've observed %for the arbitrary shape of skyrmion profile $n_z = \cos{\Lambda(r)}$.
At a small $\lambda_a$ and $ka$ (Fig.~\ref{figcr1}a) both spin-up and spin-down electrons are scattered into the same half-plane (for skyrmion orientation $\eta=+1$ it is the right half-plane).
%, it is shown in .
The increase of the skyrmion size affects the scattering in two ways making the crossover less trivial than it might be.
Firstly, at $ka>1$ higher angular harmonics with $|m|>1$ begin to contribute to the scattering amplitude and give rise to the oscillating structure of the angular dependence (Fig.\ref{figcr1}b,c).
This is a geometrical effect, a similar pattern with a predominance of forward scattering over backscattering appears
%\textcolor{red}{
in the scattering cross section
%}
%in the scattering amplitude
of a spinless particle on a cylindrical barrier, it is analogous to Mie scattering in 3D.
However, in our case the contribution of the higher angular harmonics is different for spin-up and spin-down electrons because they have different wavevectors (\ref{eqdiffk}).
This leads to the onset of the asymmetric scattering into the opposite half-plane for spin-up electrons in the range of angles close to the backscattering $\theta \sim \pi$, while spin-down electrons still scatter into the same half-plane for arbitrary magnitude of the scattering angle (Fig.\ref{figcr1}b).
%As $ka$ is further increased angular harmonics with $|m|>1$ contribute both to spin up and spin down scattering patterns which move towards small angles.
As $ka$ is further increased angular harmonics with $|m|>1$ contribute both to spin-up and spin-down scattering patterns, which move towards small angles.
%, patterns moves towards small angles, and
At that,
the asymmetry sign for spin-up and spin-down electrons is again matched in the whole range of the scattering angles (Fig.\ref{figcr1}c,d).
These peculiarities of $\Sigma_{\uparrow \uparrow}(\theta), \Sigma_{\downarrow \downarrow}(\theta)$
%landmarks
dynamics occur when neither Born series approximation is valid nor the adiabatic quasiclassical wavepacket is formed.
%The observable behavior is triggered by the interference of an electron wave at $ka \sim 2 \pi $, which is the most active at this stage.
The observable behavior arises from interference between different trajectories of a delocalized wavepacket moving in a chiral magnetization field (in this regime $ka \sim 2 \pi $ and spatial interference is most important).
%As $\lambda_a$ is further increased the weak coupling regime criteria becomes invalid, the previously mechanism providing the same assymmetry for both spins is destroyed.
As $\lambda_a$ is increased even further, the system enters the adiabatic regime with the opposite asymmetry for spin-up and spin-down electrons (Fig.\ref{figcr1}e,f).
%}
%This is the onset of the adiabatic regime with the opposite asymmetry for spin up and spin down electrons (Fig.\ref{figcr1}e,f).
%mechanism of the same
%the two pronounced peaks of opposite asymmetry merge (Fig.\ref{figcr1}e,f).

%\textcolor{red}{
The	scattering asymmetry for $\eta$-parallel channel (spin-up for $\eta=+1$) is the same at the opposite sides of the crossover:
%}
%For the $\eta$ parallel channel (spin up for $\eta=+1$) the scattering asymmetry is preserved:
spin-up is mostly scattered into the right half-plane as in the weak coupling regime. For $\eta$-antiparallel channel (spin-down for $\eta=+1$) the sign of the asymmetry is changed; spin-down electron in Fig.~\ref{figcr1}f is scattered into different (left) half-plane than for a small $\lambda_a$ (Fig.~\ref{figcr1}a).
Spin-flip scattering channels (their evolution is not presented) have very similar behavior; the difference is that at higher $\lambda_a$ they become highly suppressed in magnitude preserving the oscillating structure.
%\textcolor{red}{
%}

% подчеркнуть тезис о том, что переход не определяется не только длинной волны, но и величиной обменного взаимодействия так же
Although probably more difficult from experimental point of view, the transition from weak to adiabatic regime can be also tuned
%We would like to emphasize that a crossover to the adiabatic regime is not only tuned %by an electron wave-lenthg ($k a $ parameter), but
by the exchange constant keeping $ka=const$.
The crossover over the same range of $\lambda_a$ tuned by the exchange strength $\Delta$ %keeping $ka \gg 1$.
%Let us proceed with crossover tuned by $\Delta$ for $ka \gg 1$ -
is shown in Fig.~\ref{fig3b}.
Note, that this type of the crossover is possible only at $ka \gg 1$. If, on the opposite $ka\ll 1$ then $\lambda_a=1$ corresponds to $\Delta\gg E$ but then the spin-down electrons with a nonzero kinetic energy do not exist and so spin and charge Hall currents coincide.
As $ka\gg 1$ a number of angular harmonics contribute to the scattering already in the weak coupling  case so the whole evolution of the asymmetry difference occurs within the range of angles close to the forward scattering.
%\textcolor{red}{}

%\subsubsection{Transverse current}
\section{Charge Hall and spin Hall currents}

The topological Hall effect is measured as a current appearing in the direction perpendicular to the applied electric field. One should therefore calculate the total flux of the scattered carriers in the transverse direction.
%Given the differential cross section we calculate the quantity~\cite{prl_skyrmion}:
The corresponding quantity is the total transverse cross section ${\Sigma}_{\alpha \beta}^{tr}$
%, which
%suitable for the calculation of charge and spin currents
 given by\cite{prl_skyrmion}:
\begin{equation}
{\Sigma}_{\alpha \beta}^{tr} = \int_0^{2\pi} \Sigma_{\alpha \beta}(\theta) \sin{\theta} d\theta.
\label{eq_Sigm_Tot}
\end{equation}
An incident electron in an initial spin state $\beta$ having drift velocity $v_{\beta}$ would contribute to the transverse current $j_{\alpha \beta}$ of electrons in the final spin state $\alpha$:
\begin{equation}
j_{\alpha \beta} = 2 \pi k_{\beta} \Sigma_{\alpha \beta}^{tr}.
\label{eq_jab}
\end{equation}
The transverse currents $j_{\alpha \beta}$ have several important properties.
 %in different regimes.
%Firstly
The spin-flip channels obey $j_{\uparrow \downarrow} = j_{\downarrow \uparrow}$,
 %(in the arbitrary regime)
 which is a consequence of the hermitian property of the Hamiltonian (\ref{eq_Pot}).
In the adiabatic regime the spin-flip channels are suppressed $j_{\uparrow \downarrow} = 0$, while transverse spin-conserving currents
%channels
have the same magnitude and the opposite sign for spin-up and spin-down $j_{\uparrow \uparrow} = - j_{\downarrow \downarrow}$. This is in full accordance with the
Berry curvature having opposite sign for spin-up and spin-down carriers.
%because an effective magnetic field due to Berry's phase differs only by sign.
%Regarding the adiabatic regime when spin-flip channels are suppressed and spin up and down electrons move in having opposite sign .
%We found that at this stage the transverse currents (Eq.~\ref{eq_jab}) within spin conserving channels have equivalent absolute values and opposite signs $j_{\uparrow \uparrow} = - j_{\downarrow \downarrow}$.
In the weak coupling regime when both $\lambda_a\ll1$ and $\Delta/2E \rightarrow 0$
the asymmetric scattering does not depend on spin, the currents of spin-conserving channels coincide $j_{\uparrow \uparrow} = j_{\downarrow \downarrow}$, while the spin-flip transverse currents are two times greater $j_{\downarrow \uparrow} = 2 j_{\uparrow \uparrow}$.

For an unpolarized incident electron flux
the transverse charge current $j_H$ characterizing the charge Hall effect and the spin current $j_{SH}$ characterizing the spin Hall effect can be calculated as:

\begin{equation*}
\begin{aligned}
& j_H = j_{\uparrow \uparrow} + j_{\downarrow \downarrow} + j_{\uparrow \downarrow} + j_{\downarrow \uparrow},
\\
& j_{SH} = j_{\uparrow \uparrow} - j_{\downarrow \downarrow} + j_{\uparrow \downarrow} -  j_{\downarrow \uparrow}.
\end{aligned}
\end{equation*}

\begin{figure*}
	\centering	
	\includegraphics[width=1.\textwidth]{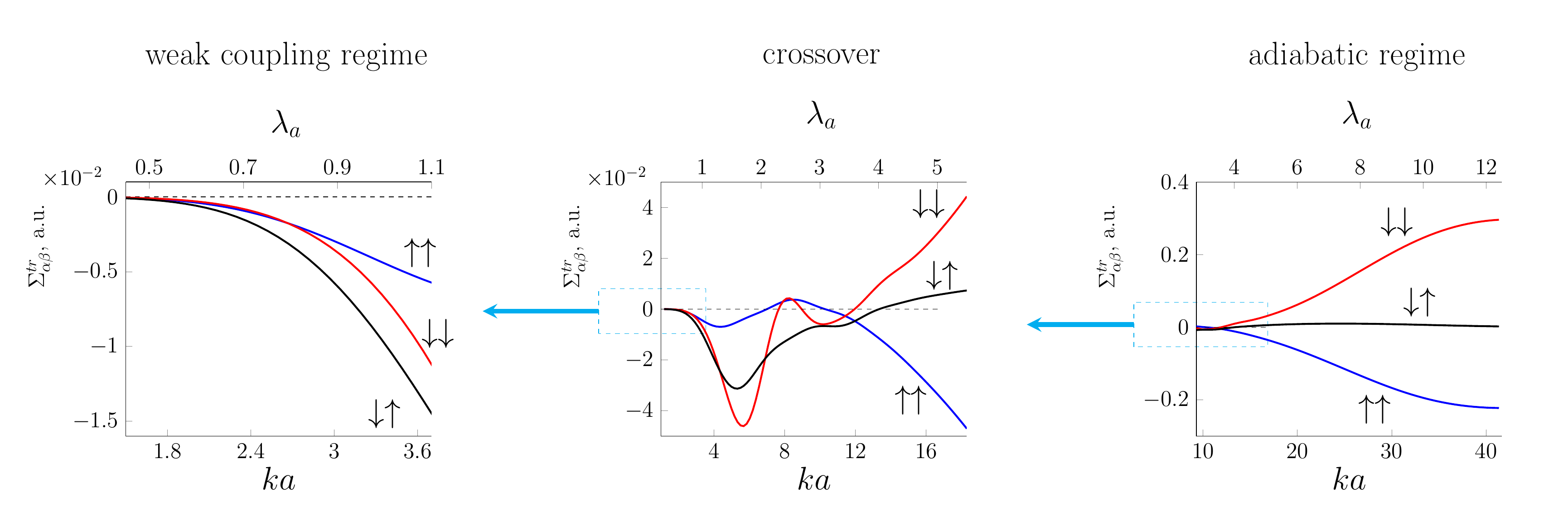}%{fig-disc-1.eps}
	%{fig-1.pdf}%{fig-1.png}%{3-3.jpg}%{BDS.eps}
	%\includegraphics[width=0.55\textwidth]
	\caption{
		Evolution of total asymmetric flux $\displaystyle \Sigma_{\alpha \beta}^{tr} = \int\limits_0^{2\pi} \Sigma_{\alpha \beta}(\theta) \sin{\theta} d\theta $ with skyrmion size for an electron scattering on a magnetic skyrmion with $Q=+1$, $\kappa=+1$. The exchange splitting is fixed $\Delta/2E = 0.3$. $\Sigma_{\alpha \beta}^{tr}$ is given in units of skyrmion radius $a/2$.				
	}
	\label{figdisc1}
\end{figure*}

		\begin{figure*}
			\centering	
			\includegraphics[width=1.\textwidth]{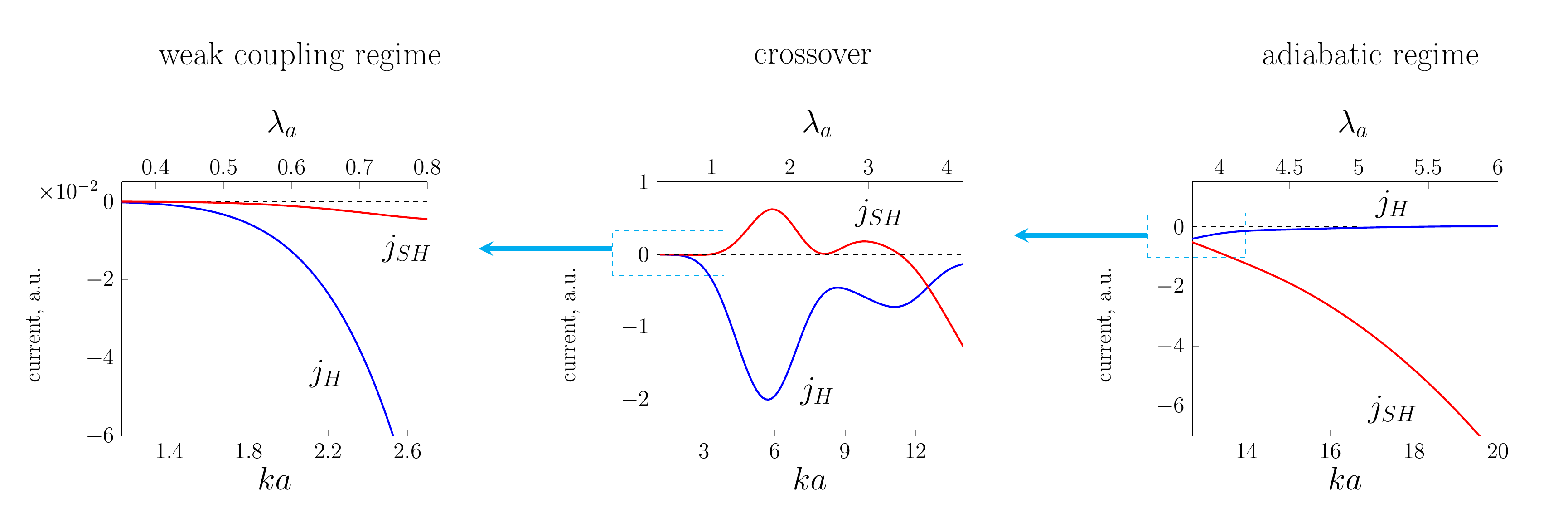}%{fig-disc-1.eps}
			%{fig-1.pdf}%{fig-1.png}%{3-3.jpg}%{BDS.eps}
			%\includegraphics[width=0.55\textwidth]
			\caption{Crossover between transverse charge current $j_H$ at small $\lambda_a$ and spin Hall current $j_{SH}$  at large $\lambda_a$ driven by variation of skyrmion size $a$. The exchange splitting is fixed $\Delta/2E = 0.3$.
			}
			\label{figHall1}
		\end{figure*}	
		
		\begin{figure*}
			\centering	
			\includegraphics[width=1.\textwidth]{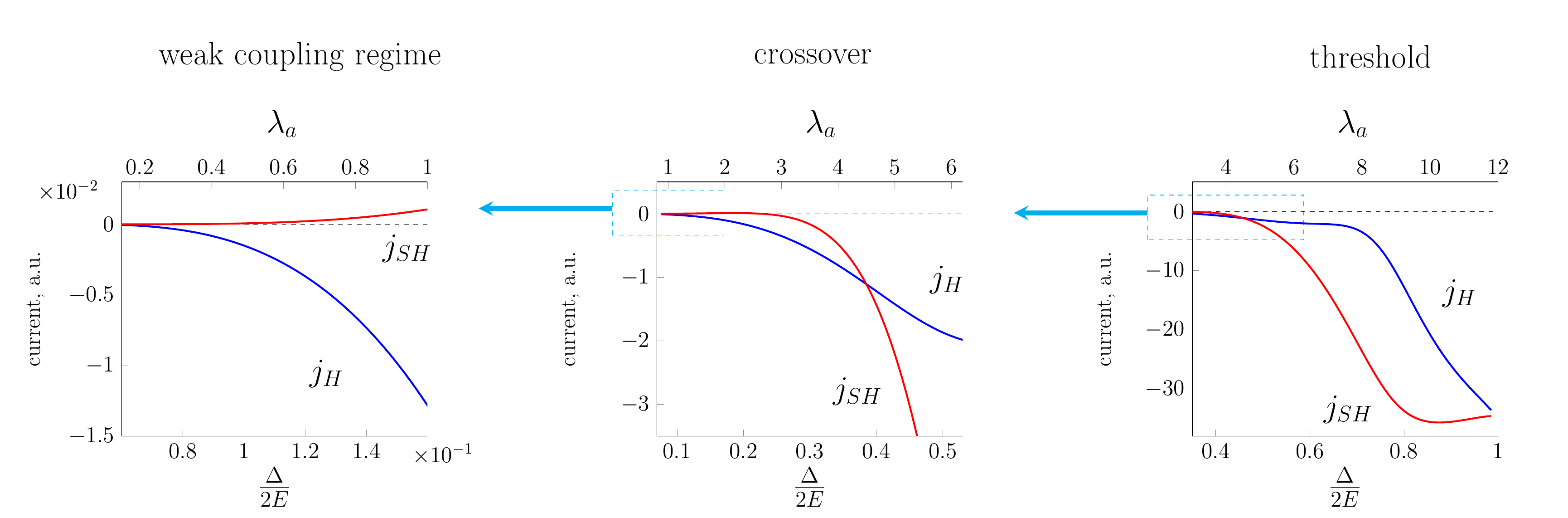}%{fig-disc-1.eps}
			%{fig-1.pdf}%{fig-1.png}%{3-3.jpg}%{BDS.eps}
			%\includegraphics[width=0.55\textwidth]
			\caption{Evolution of transverse charge $j_H$ and spin $j_{SH}$ currents with exchange band splitting $\Delta$. The skyrmion size is fixed $ka = 12$.
			}
			\label{figHall2}
		\end{figure*}

For the crossover driven by variation of a skyrmion size the evolution of ${\Sigma}_{\alpha \beta}^{tr}$ for
different scattering channels
% spin-conserving and spin-flip scattering channels
is shown in Fig.~\ref{figdisc1} (for the same set of parameters as in Fig.~\ref{figcr1}).
The corresponding charge current $j_H$ and spin current $j_{SH}$ evolution is presented in Fig.~\ref{figHall1}.

%The evolution of ${\Sigma}_{\alpha \beta}^{tr}$ tuned by a skyrmion size $a$ for %different scattering channels is shown in Fig.~\ref{figdisc1}
%Left and right panels in Fig.~\ref{figdisc1} correspond simultaneously to the weak %coupling and adiabatic limiting regimes.
At a small $\lambda_a$ (Fig.~\ref{figdisc1}, left panel)
the asymmetrical transverse electron flux in each scattering channel (spin conserving and spin-flip) is of the same sign, that is the carriers are preferably scattered to the same transverse direction regardless their spin.
The charge Hall current therefore strongly prevails over the spin Hall current as clearly seen in Fig.~\ref{figHall1}.
%scattering produces
%electron within each channel is preferably scattered to the same half-plane.
At a large $\lambda_a \gtrsim 5$ spin-flip channels are suppressed; spin-up and spin-down electron are scattered in the opposite directions (right panel Fig.~\ref{figdisc1}). Consequently, spin Hall current strongly dominates over the charge Hall current (Fig.~\ref{figHall1}, right panel).
%the contribution of spin-flip scattering is suppressed, the spin conserving channels %having the opposite scattering asymmetry contribute to the transverse spin current %(Fig.~\ref{figdisc1}).
In this regime similarly to AHE the charge Hall current can appear only if the incident electrons are spin polarized, i.e. there is unequal number of spin-up and spin-down electrons.
It is worthwhile noticing that for a fixed wavelength the magnitude of
the cross section increases with the
skyrmion size. 	
Let us discuss a crossover tuned by a skyrmion size in more details. The main feature of the intermediate region is that the asymmetric total cross section $\Sigma_{\alpha \beta}^{tr}$ has a nontrivial oscillating structure with pronounced peaks.
% within each scattering channel.
These oscillations reflect the complex pattern of $\Sigma_{\alpha \beta}(\theta)$:
% evolution of these peaks correlates with landmarks that
The first peak of $\Sigma_{\alpha \beta}^{tr}$ occurs when spin-up and spin-down channels start to diverge (Fig.~\ref{figcr1}b),
%\textcolor{red}{
while the second peak emerges when spin-down and spin-up scattering channels
restore their similar behavior (see Fig.~\ref{figcr1}c,d).
Hence, the oscillating structure of $\Sigma_{\alpha \beta}^{tr}$ reflects
%is related to the same physical processes that affect the pattern of $\Sigma_{\alpha %\beta}(\theta)$ in Fig.~\ref{figcr1}; namely the oscillations are relevant to an
the electron wave interference at $ka \sim 2 \pi$ and deviation from Born series towards adiabatic scenario.
%is triggered by an electron wave interference at $ka \sim 2 \pi$ and deviation from Born series towards adiabatic scenario.
%}
%Let us discuss a crossover in more details. The main feature of the intermediate region is that the asymmetric total flux $\Sigma$ has an oscillating structure, with a very pronounced peaks within each scattering channels. This structure emerges appears following the oscillating evolution of $\Sigma$ on $\theta$. Therefore the discovered oscillations of $\Sigma^{tr}$ originate from an electron wave interference at $ka \sim 1$ and deviation from Born approximation towards adiabatic scattering scenario. The nontrivial structure of $\Sigma_{\alpha \beta}^{tr}$ leads to charge $j_H$ and spin $j_{SH}$ transverse currents.

%\textcolor{red}{
These properties of the transverse flux of the carriers lead to a nontrivial behavior of transverse charge $j_H$ and spin $j_{SH}$ currents shown in Fig.~\ref{figHall1}.
%}
The discovered oscillating structure of the transverse currents
%We would like to note that the discovered oscillating structure of the THE crossover
can be used to differentiate THE contribution from other Hall contributions. We predict that upon varying skyrmion size or Fermi level the observable topological Hall response would acquire a nonmonotonic and oscillating structure, which can be regarded as a characteristic feature of THE response when treating the experiments.
%\textcolor{red}{From the analysis given above we conclude that even for $\lambda_a>1$ using the adiabatic approach can lead to qualitatively wrong results when estimating the topological Hall effect if the adiabatic parameter value falls in the range of the crossover.}

%However the crossover that we observe has a nontrivial character.
%The main feature of the intermedaite region is presence of oscillations and peaks.
%We attribute this behaviour to an electron wave interference on a skyrmion scale.
%The point is that when $\lambda_a$ falls into the intermediate region, a spacial %electron phase $i {\bf kr}$ aquired on a skyrmion size is of order of one; so the %interference pecularities are essential at this stage. The interference manifests %itself as an appearance of peaks upon changing a skyrmion size (tuning the crossover by %an exchange constant $\Delta$ doesn't naturally exhibit the discussed behaviour, see %Fig.~\ref{figHall2}).

The evolution of $j_H, j_{SH}$ driven by variation of the exchange strength $\Delta$ is shown in Fig.~\ref{figHall2}.
Unlike the previously considered case there is no oscillating structure associated with the geometrical interference effect as the skyrmion size is not changed
(the parameter $ka$ is fixed). An important difference from the skyrmion size driven crossover is that above the threshold $\Delta/2E = 1$ there is only one spin subband with a nonzero kinetic energy
(spin-up for the skyrmion orientation $\eta = +1$) and hence the transverse charge and spin currents coincide.
At a small $\lambda_a$ the charge current $j_H$ dominates over $j_{SH}$.
%\textcolor{red}{this is required by symmetry arguments valid in the perturbative %region.}
% as in the skyrmion size driven crossover.
For larger $\Delta$ both currents become comparable, with further increase of $\lambda_a$
there is no suppression of charge current $j_H$, instead the scattering in the spin-down channel is being suppressed so that the contribution to both charge and spin currents comes from the spin-up channel finally leading to $j_H=j_{SH}$ (Fig.~\ref{figHall2}).

\begin{figure}
\includegraphics[width=.3\textwidth]{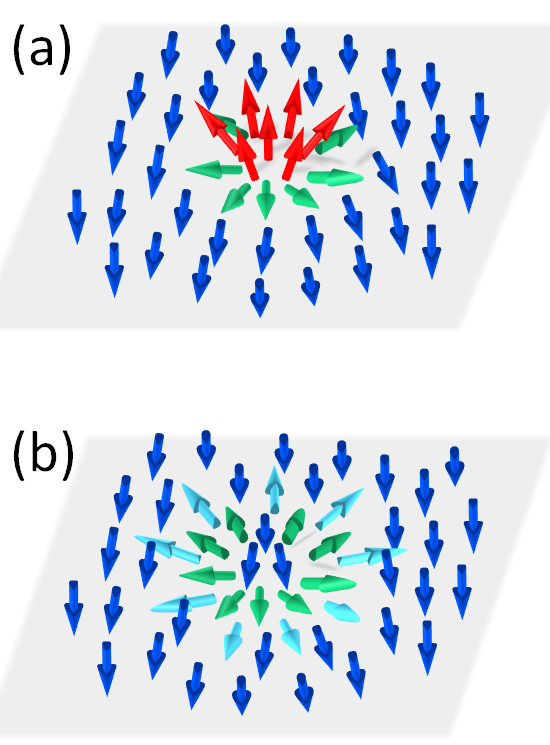}%{fig-vortex.jpg}%{pict_4-2.pdf}%{fig2.pdf}%{pict_4_hor.pdf}%{fig-cr-2.eps}%{fig-1.pdf}%{fig-1.png}%{3-3.jpg}%{BDS.eps}
	\caption{
		{
		Two types of chiral magnetic vortices: (a) - magnetic skyrmion with a nonzero topological charge; (b) - magnetic vortex with trivial topology. Both structures have nonzero vorticity $\kappa = +1$ and produce Hall response.
		}
	}	
	\label{figVortex}
\end{figure}
\begin{figure*}
	\centering	
	\includegraphics[width=1.\textwidth]{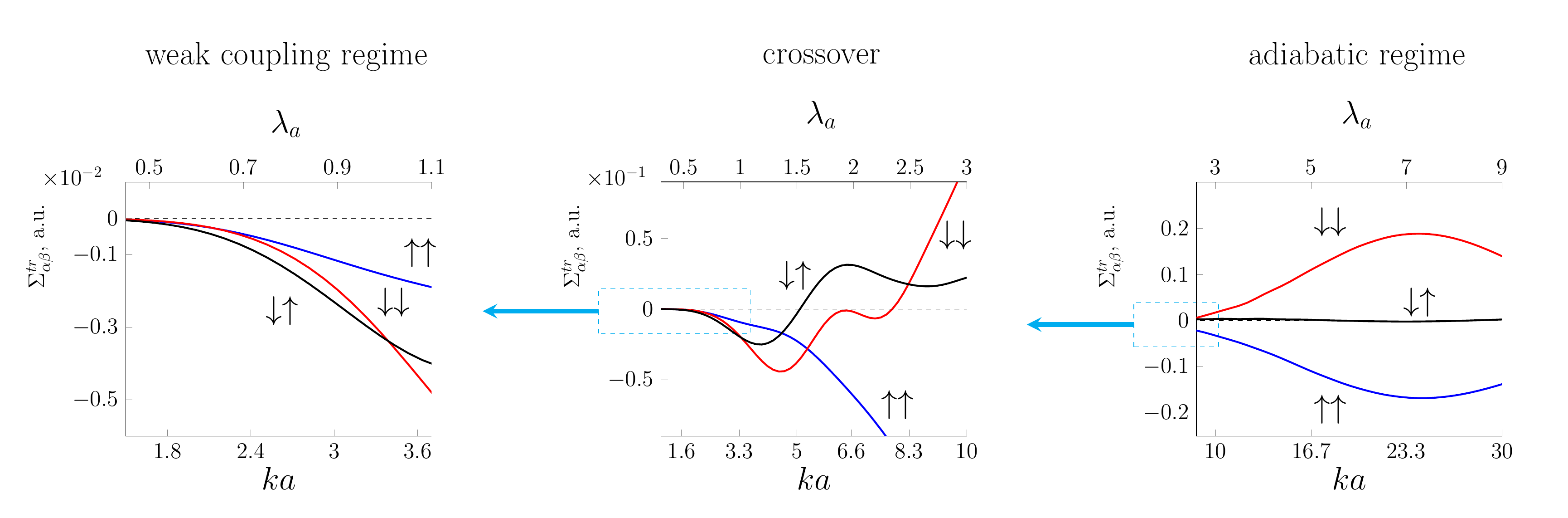}%{fig-disc-1.eps}
	%{fig-1.pdf}%{fig-1.png}%{3-3.jpg}%{BDS.eps}
	%\includegraphics[width=0.55\textwidth]
	\caption{
		Evolution of total asymmetric flux $\displaystyle \Sigma_{\alpha \beta}^{tr}$ with skyrmion size for an electron scattering on a topologically trivial chiral magnetic vortex with $Q =0$ and $\kappa = +1$. The exchange splitting is fixed $\Delta/2E = 0.3$. $\Sigma_{\alpha \beta}^{tr}$ is given in units of vortex radius $a/2$.				
	}
	\label{figQ}
\end{figure*}
\section{Scattering on a topologically trivial magnetic vortex}
\label{secTriv}
Another interesting finding of our study is that
%contrary to the previously claimed
%finding of
%It's wothwhile noticing that
%Another interesting finding
the magnetic vortex topological charge itself is not essential for the discussed phenomena.
Our theory predicts that even chiral configurations with zero topological charge such as a co-vortice shown in Fig.~\ref{figVortex}b
can exhibit transverse scattering properties similar to that of a topologically charged magnetic skyrmion (Fig.~\ref{figVortex}a).
%Furthermore, all mentioned behaviour of asymmetric scattering and the existane of crossover are preserved (See Fig.\ref{figdisc1}).
%This finding has been missed in the mean field theories since they considered only global (topological) properties of the magnetization field ignoring its local chiral fine structure.
In Fig.~\ref{figQ} we present evolution of the total transverse scattering cross section $\Sigma_{\alpha \beta}^{tr}$ for the topologicaly trivial magnetic vortex.
Inside its core $r<a/2$ the vortex is parametrized with the profile $\Lambda(r) = 4 \pi r/a (1-2 r/a)$.
%\textcolor{red}{
Similarly to the magnetic skyrmion considered in the previous sections, there are different regimes of asymmetric electron scattering.
At a small $\lambda_a$ (Fig.~\ref{figQ}, left panel) the charge transverse effect dominates with each scattering channel having the same sign of $\Sigma_{\alpha \beta}^{tr}$. At a large $\lambda_a$ (Fig.~\ref{figQ}, right panel) there is a pronounced spin Hall effect with spin-flip channels being suppressed.
The intermediate region $\lambda_a \sim 1$ exhibits an oscillating crossover.

This finding highlights that the microscopic origin of THE originates from a local chiral ordering of the magnetic moments rather than from a global topology of the magnetization field. THE can be expressed in terms of a topological characteristic of the magnetic structure only when the mean field approximation is applicable
%is it is possible to describe an electron interaction with noncollinear area of %magnetization as an electron motion in some averaged field
and the local deviations of the magnetization can be neglected\cite{Ye1999,BrunoDugaev}.
%\textcolor{red}{
The mean field approach is adequate for arrays of magnetic skyrmions\cite{Zhou-1,Zhou-2}, skyrmion lattices\cite{MnSiAPhase,FeGe_THE} and other dense skyrmion systems.
%averaged permanent chiral field\cite{Ye1999,BrunoDugaev}, i.e. via meanfield approximation ignoring local deviations from the averaged value.
However, an electron scattering on an individual chiral vortex cannot be reduced to an electron motion in a homogeneous effective magnetic field.
{
\section{Summary}
}
%In conclusion we consider the Topological Hall effect in various regimes.
The presented analysis of microscopic electron scattering on a chiral magnetization field
enabled us to formulate the following features of the topological Hall effect.
%Based on the microscopic model of an electron scattering on a single magnetic skyrmion we succeed in investigating the %THE behavior for the arbitrary values of the adiabatic parameter.
When both free carriers spin subbands are involved there are two qualitatively different regimes characterized by the adiabatic parameter $\lambda_a$.
%activated we indicate the existance of two limiting regimes with qualitatevely different behavior.
In the range $\lambda_a \ll 1$ a charge carrier exchange interaction with a skyrmion leads to the transverse charge current with a negligible spin Hall effect. On the contrary, in the adiabatic regime $\lambda_a \gg 1$ the spin Hall effect dominates and the transverse charge current appears only if there is a substantial spin polarization of the carriers, this regime is similar to the anomalous Hall effect.
Our theory allowed us to trace the nontrivial crossover between the two regimes for the intermediate values of $\lambda_a$.
For the most realistic crossover driven by a skyrmion size or the carriers Fermi level the
transverse spin and charge currents oscillate with $\lambda_a$ reflecting the
%Hence, the oscillating structure of $\Sigma_{\alpha \beta}^{tr}$ reflects
%is related to the same physical processes that affect the pattern of $\Sigma_{\alpha %\beta}(\theta)$ in Fig.~\ref{figcr1}; namely the oscillations are relevant to an
electron wave interference at $ka \sim 2 \pi$ and deviation from the second Born approximation towards adiabatic scenario. The discovered characteristic feature of topological Hall effect can be used as a new tool for experimental detection of THE.

%\textcolor{red}{The behaviour within the crossover region.}
%Thus, the apparent contradiction between the results of adiabatic and perturbative theoretical approaches to THE have been eliminated for the first time.
%We discovered the existence of a THE nontrivial crossover in the intermediate region of $\lambda_a$.

\section{Acknowledgments}
We thank M.~M.~Glazov for helpful discussion.
The work
%The research and development of the theoretical model
has been carried out under the financial support
%of a
%Grant
from Russian Science Foundation,
project 17-12-01182 (analytical theory)%
%(physical model) 
 and
project 17-12-01265 (numerical calculations),
%(analytical calculations),
K.S.D. thanks Russian Foundation of Basic Research  (Project no.16-32-00798)
for financial support.
% on numerical calculations.

%via the multichannel variable
\appendix
\section{Methods}
\label{AppMethods}
%\textcolor{blue}{

In this Appendix we consider the calculation of scattering amplitude ${f}_{\alpha \beta}(\theta)$ via the phase-function method.
The scattering potential of a chiral magnetic vortex (\ref{eq_Pot}) commutes with operator $ \hat{j} = - i \partial_{\theta} + \kappa \hat{S}_z$, so the eigenfunctions are characterised by the corresponding quantum number $j$ having half-integer values.
%However,
Throughout the paper we label these eigenstates by the angular momentum projection $m = j - 1/2$ (taking integer values). The eigenfunctions are written:
\begin{equation}
\label{eqphim}
\psi _m  = {e^{im\theta }}
\begin{pmatrix}
g_m^{(1)}(r) \\
e^{i\kappa \theta + i \gamma} g_{m}^{(2)}(r)
\end{pmatrix},
\end{equation}
where $g_m^{(1,2)}(r)$ are the functions of radius vector $r$; they satisfy
%In order to calculate $f_{\alpha \beta}$ we have to find the exact $\psi_m$. Substituting (\ref{eqphim}) into (\ref{eq_Ham}) we get for each $m$
the system of equations:
%only on the radius vector magnitude $r$:
\begin{equation}\label{eq_wf_equations}
\left({H}_m^{'}  + \frac{\Delta}{2E} k^2 {W}\right) {q}_m= 0,
\end{equation}
where ${q}_m \equiv (g_m^{(1)}(r), g_m^{(2)}(r))^{T}$ is the two-component function of $r$, %$\gamma = m_{\ast} \Delta/\hbar^2$,
the matrices ${H}_m^{'}$, ${W}$ are given by:
\begin{equation*}%\label{eq_Hm}
%\begin{aligned}
{H}_m^{'} = \begin{pmatrix}
\frac{1}{r} \partial_r \left(r \partial_r \right)   - \frac{m^2}{r^2} + k_{\uparrow}^2& 0
\\
0 & \frac{1}{r} \partial_r \left(r \partial_r \right)  - \frac{(m+\kappa)^2}{r^2} + k_{\downarrow}^2
\end{pmatrix},
\end{equation*}
\begin{equation}\label{eq_W}
{W} =
\begin{pmatrix}
- \eta (1-n_z(r)) & n_{\parallel}(r)
\\
n_{\parallel}(r) & \eta (1-n_z(r))
\end{pmatrix}.
%\end{aligned}
\end{equation}
The operator ${H}_m^{'}$ corresponds to the free motion Hamiltonian, while ${W}$ is the perturbation due to the magnetic vortex.
Outside of the core $r>a/2$ the term ${W}$ vanishes so that ${q}_m$ becomes a combination of Bessel functions:
%, in the absence of the scattering potential there are two independent solutions corresponding to unmixed spin-up and spin-down states:
%when scattering potential is abscent
\begin{equation*}
\label{eq_Kmat}
q_m^{(1)} =
\begin{pmatrix}
J_{m}(k_{\uparrow} r) - K_m^{11} Y_{m}(k_{\uparrow} r)
\\
- K_m^{21} Y_{m+\kappa}(k_{\downarrow} r)
\end{pmatrix},
\end{equation*}
\begin{equation*}
\label{eq_Kmat2}
q_m^{(2)} =
\begin{pmatrix}
- K_m^{12} Y_{m}(k_{\uparrow} r)
\\
J_{m+\kappa}(k_{\downarrow} r) - K_m^{22} Y_{m+\kappa}(k_{\downarrow} r)
\end{pmatrix},
\end{equation*}
where $J_m,Y_m$ are the $m$-th order Bessel functions of the first and second kind, respectively. $K_m$  is $2\times 2$ constant matrix which determines the $m$-th scattering matix $S_m = (1+iK_m)(1-iK_m)^{-1}$. The scattering amplitude $f_{\alpha \beta}(\theta)$ is given by a sum over elements of $S_m$ (Eq.~\ref{eq_f}).
%(that are included into the scattering amplitude $f_{\alpha \beta}(\theta)$ via Eq.~\ref{eq_f}).
The matrices $K_m$ (or $S_m$) are found from the exact solution of Eq.~(\ref{eq_wf_equations}) inside the skyrmion core $r<a/2$.
%calculated using phase-functions method\cite{Babikov1} from the exact solution of Eq.~(\ref{eq_wf_equations}).
%whose coeffitients and phase shifts will determine the scattering amplitude $f_{\alpha \beta}$.

The partial scattering matrices ${S}_m$ were calculated using phase-functions method\cite{Babikov1}. This technique considers the scattering parameters ($K_m(r), S_m(r)$) as functions of radius vector $r$ in the region $r<a/2$.
The matrices $S_m(r)$ describe the scattering on a potential being cut off at the point $r$. The advantage of the approach is that instead of solving Schr\"odinger equation (\ref{eq_Ham}) one should solve numerically the first order matrix differential nonlinear equation for $S_m(r)$:
\begin{equation}
\label{eq_Ap1}
\frac{d {S}_m}{d r} = i \frac{\pi r}{4} \frac{\Delta}{2E} k^2 \left( {R}_m^{-} + {S}_m {R}_m^{+}  \right) {W}  \left( {R}_m^{-} +  {R}_m^{+}{S}_m  \right),
\end{equation}
	where
	\begin{equation*}
	\begin{aligned}
	& {R}_m^{\pm}(r) = \begin{pmatrix}
	H_m^{(1,2)}(k_{\uparrow}  r) &  0
	\\
	0 & H_{m+\kappa}^{(1,2)}(k_{\downarrow}  r)
%	J_m(k_{\uparrow}  r) \pm i Y_m(k_{\uparrow}  r) &  0
%	\\
%	0 & J_{m+\kappa}(k_{\downarrow}  r) \pm i Y_{m+\kappa}(k_{\downarrow}  r)
	\end{pmatrix},
	\end{aligned}
	\end{equation*}
%\end{widetext}
%with the obvious
$H_m^{(1,2)}$ are the $m$-th order Hankel functions of the first and second kind,
and $W(r)$ is the potential defined by the vortex structure (\ref{eq_W}). The bondary condition is $S_m^{\alpha \beta}(r=0) = \delta_{\alpha \beta}$. Since we consider a vortex of a finite size with uniform background magnetization outside,
%of it $r>a/2$,
%(there is no perturbation over the host magnetization outside its core $r>a/2$),
the value of the partial scattering matrix at the boundary $S_m(a/2)$ gives the scattering amplitude (\ref{eq_f}).

%determines the correct scattering matrix.
%}

\section{2D scattering on a triad of spins}
\label{AppTriad}
%highlight
In this Appendix we explain the microscopic origin of the charge Hall response due to non-zero spin chirality in the weak-coupling regime.
 %when spin-flip processes are fully activated.
 Let us consider a scattering of an electron on a triad of non-coplanar spins (Fig.~\ref{figtriada}). The incident electron comes along $x$-direction and the scatterers forming the triad are located symmetrically with respect to the reflection in $xz$ plane so that their spatial arrangement does not produce any scattering asymmetry in transversal $y$ direction.
We take the scattering potential in the form:
\begin{equation}
V =  \sum\limits_{i = 1,2,3} {V_i},\,\,\,\,\,\,\,\,\,\,   V_i =  - A{{{\bf{n}}_i}{{\boldsymbol \sigma }}}{\delta \left( {{\bf{r}} - {{\bf{r}}_i}} \right)},
\label{eq_V_app}
\end{equation}
where ${\bf r_i}$ is the radius-vector of the $i$-th scatterer,   ${\boldsymbol{\sigma }}$ is the vector of Pauli matrices,
${{\bf n}_i}$ are the  unit length vectors indicating magnetization directions of the three scatterers,
the constant $A$ accumulates the exchange interaction strength and the magnitudes of the electron and the scatterers spins.

In the weak coupling case the kinetic energy difference between spin-up and spin-down electrons can be neglected so that $k_\uparrow\approx k_\downarrow \equiv k$.
The spin-dependent scattering amplitude $f_{\alpha\beta}(\theta)$ for scattering from a state $\left|{\bf k},\beta\right\rangle$
(${\bf k}$ points along $x$-direction)
into a state $\left|{\bf k'},\alpha\right\rangle$ defined in (\ref{eq_PSI})
can be expressed in terms of $T$-matrix as\cite{Adhikari}:
\begin{equation}
{f_{\alpha \beta }}\left( \theta  \right) =  - \frac{{{m_*}}}{{{\hbar ^2}}}\frac{{{e^{i\pi /4}}}}{{\sqrt {2\pi k} }}\langle {\bf{k'}},\alpha |T\left| {{\bf{k}},\beta } \right\rangle,
\label{eq_f_T}
\end{equation}
where $\theta$ is the scattering angle between ${\bf k}$ and ${\bf k'}$,
$\alpha,\beta$ are spin indices.
In the second Born approximation the $T$-matrix is given by:
\[T = V + V{G_0}V,\]
where $V$ is the scattering potential (Eq.~\ref{eq_V_app}), $G_0$ is the 2D Green's function of a free propagating electron.
%single

To calculate $f_{\alpha \beta}$ via Eq.~(\ref{eq_f_T}) we need the matrix elements of $T$-matrix:
\begin{align}
&\langle {\bf{k'}}, \alpha |T\left| {{\bf{k}}, \beta} \right\rangle  = \langle {\bf{k'}}, \alpha|V\left| {{\bf{k}}, \beta } \right\rangle  + \langle {\bf{k'}}, \alpha |V{G_0}V\left| {{\bf{k}}, \beta} \right\rangle, \\
&\langle {\bf{k'}},\alpha |V\left| {{\bf{k}},\beta } \right\rangle  =  - A\sum\limits_{i = 1,2,3} {{e^{i\left( {{\bf{k}} - {\bf{k'}}} \right){{\bf{r}}_i}}}} {v_{i,\alpha \beta }},\\
&\langle {\bf{k'}},\alpha |V{G_0}V\left| {{\bf{k}},\beta } \right\rangle  = \nonumber\\ &=  - i\frac{{{A^2}{\pi ^2}m}}{{{\hbar ^2}}}\sum\limits_{\alpha '} {\sum\limits_{i,j} {{e^{i\left( {{\bf{k}}{{\bf{r}}_j} - {\bf{k'}}{{\bf{r}}_i}} \right)}}} {H_0^{(1)}}\left( {{k}{r_{ij}}} \right){v_{i,\alpha \alpha '}}{v_{j,\alpha '\beta }}},
\end{align}
where ${{\bf{r}}_{ij}} = {{\bf{r}}_i} - {{\bf{r}}_j}$, ${r_{ij}} = \left| {{{\bf{r}}_{ij}}} \right|$,
$H_0^{(1)}$ is  Hankel function of the first kind,
and $v_{i,\alpha \beta}$ are spin matrix elements:
\begin{equation}
{v_{i,\alpha \beta }} = \left\langle \alpha  \right|{{\bf{n}}_i}\boldsymbol{\sigma} \left| \beta  \right\rangle
= {\left( {\begin{array}{*{20}{c}}
		{{n_{iz}}}&{{n_{i-}}}\\
		{{n_{i+}} }&{ - {n_{iz}}}
		\end{array}} \right)}_{\alpha \beta}.
	\label{eqvij}
\end{equation}
The scattering cross section
%for small exchange interaction
is given by $d \sigma_{\alpha \beta}/d\theta = |f_{\alpha \beta}|^2$, so we calculate
\begin{align}
&{\left| {\langle {\bf{k'}},\alpha |T\left| {{\bf{k}},\beta } \right\rangle } \right|^2} = {A^2}\sum\limits_{i,j} {{e^{i\left( {{\bf{k}} - {\bf{k'}}} \right){{\bf{r}}_{ij}}}}} {v_{i,\alpha \beta }}v_{j,\alpha \beta }^* + \nonumber\\
 &\frac{{{2A^3}{\pi ^2}m}}{{{\hbar ^2}}}  {\mathop{\rm Im}\nolimits}  	\sum\limits_{i,j,l,\alpha'} {{e^{i\left( {{\bf{k}}{{\bf{r}}_{jl}} + {{\bf{k}}^\prime }{{\bf{r}}_{li}}} \right)}}}
%	\sum\limits_{\alpha '}
{{H_0^{(1)}}\left( {{k_{\alpha '}}{r_{ij}}} \right){v_{i,\alpha \alpha '}}{v_{j,\alpha '\beta }}v_{l,\alpha \beta }^*}  \nonumber\\
&+ O\left( {{A^4}} \right).
\label{eqT2}
\end{align}

We focus on the scattering asymmetry related to the spin chirality of the magnetization field, for the considered triad that is $\chi_c = {\bf n}_1 \cdot \bigl[{\bf n}_2 \times {\bf n}_3\bigr]$.
Thus, any chirality related phenomena would include all three spins of the triad.
 %it appears in the third order in $A$.
The first term in (\ref{eqT2}) corresponding to the first Born approximation
consists of combinations of only two spin matrix elements (\ref{eqvij})
 and appears to be irrelevant to the chirality related scattering.
%is not relevant since it is a combination of 2 spins.
The spin chirality $\chi_c$ first appears in the third order on the exchange interaction $A$.
%in (\ref{eqT2}
Let us consider the third order terms for spin conserving scattering channel $(\alpha=\beta)$. We obtain:
%Then for the asymmetrical spin conserving scattering from (\ref{eqMatrEl}) we keep only third order terms:
%with $i\ne j \ne l$:
%leave the chirality aware terms:
\[{\left| {{T_{{\bf k'k}\alpha \alpha }}} \right|^2}_{{A^3}} = \frac{{{A^3}2{\pi ^2}m}}{{{\hbar ^2}}}{\mathop{\rm Im}\nolimits}  {\sum\limits_{i, j, l} {{e^{i\left( {{\bf{k}}{{\bf{r}}_{jl}} + {\bf{k}}'{{\bf{r}}_{li}}} \right)}}} {H_0^{(1)}}\left( {k{r_{ij}}} \right){\Omega _{ijl,\alpha }}}, \] where
\begin{align}
%&{\Omega _{ijl, \uparrow  \uparrow }} = {n_{lz}}\left( {{{\bf{n}}_i}{{\bf{n}}_j} + %i{{\left[ {{{\bf{n}}_i} \times {{\bf{n}}_j}} \right]}_z}} \right)\nonumber \\
&{\Omega _{ijl, \uparrow  \uparrow }} = {n_{lz}}{{\bf{n}}_i}{{\bf{n}}_j} + i\chi _z^{lij} \nonumber \\
&{\Omega _{ijl, \downarrow  \downarrow }} =  - {n_{lz}}{{\bf{n}}_i}{{\bf{n}}_j} + i\chi _z^{lij}.
%&{\Omega _{ijl, \downarrow  \downarrow }} =  - {n_{lz}}\left( {{{\bf{n}}_i}{{\bf{n}}_j} %- i{{\left[ {{{\bf{n}}_i} \times {{\bf{n}}_j}} \right]}_z}} \right)
\label{eqnuchiral}
\end{align}
The
%terms contain the
chirality aware term here is the imaginary part:
\[ \chi_{q}^{lij} = n_{l q} \left[ {{{\bf{n}}_i} \times {{\bf{n}}_j}} \right]_q
	\hspace{1cm} q={x,y,z}.\]
%last term in brackets, it contains all three projecions of the components of the %magnetization.
As it is clearly seen in (\ref{eqnuchiral}) this part appears to have the same sign for spin-up and spin-down scattering channels.
Analogously, for the spin-flip third order scattering channels we obtain:
\begin{align}
\label{eqOmegaflip}
&{\Omega _{ijl, \uparrow  \downarrow }} = {\left[ {{{\bf{n}}_l} \times \left[ {{{\bf{n}}_i} \times {{\bf{n}}_j}} \right]} \right]_z} + i\left( {\chi _x^{lij} + \chi _y^{lij}} \right)\nonumber \\
&{{\Omega _{ijl, \downarrow  \uparrow }} =  - {{\left[ {{{\bf{n}}_l} \times \left[ {{{\bf{n}}_i} \times {{\bf{n}}_j}} \right]} \right]}_z} + i\left( {\chi _x^{lij} + \chi _y^{lij}} \right).}
%{\Omega _{ijl, \uparrow  \downarrow }} &= {\left[ {{{\bf{n}}_l} \times \left[ %{{{\bf{n}}_i} \times {{\bf{n}}_j}} \right]} \right]_z} \nonumber\\& + i\left( %{{n_{lx}}{{\left[ {{{\bf{n}}_i} \times {{\bf{n}}_j}} \right]}_x} + {n_{ly}}{{\left[ %{{{\bf{n}}_i} \times {{\bf{n}}_j}} \right]}_y}} \right)\nonumber \\
%{\Omega _{ijl, \downarrow  \uparrow }} &=  - {\left[ {{{\bf{n}}_l} \times \left[ %{{{\bf{n}}_i} \times {{\bf{n}}_j}} \right]} \right]_z} \nonumber\\ & + i\left( %{{n_{lx}}{{\left[ {{{\bf{n}}_i} \times {{\bf{n}}_j}} \right]}_x} + {n_{ly}}{{\left[ %{{{\bf{n}}_i} \times {{\bf{n}}_j}} \right]}_y}} \right)
\end{align}
Similarly to the diagonal channels (\ref{eqnuchiral}),
%spin-flip terms contain different combination of three spins ${\bf n}_i$. However
the sign of the spin-chirality terms is also the same for the two opposite spin-flip scattering channels (\ref{eqOmegaflip}).
Finally, the spin chirality aware part contributing to the cross section appears to be the following:
\begin{widetext}
	\begin{align}
{\left| {{T_{{\bf{k'k}}\alpha \beta }}} \right|^2_{\chi}} = \frac{{{A^3}2{\pi ^2}m}}{{{\hbar ^2}}}{\rm{Re}}\sum\limits_{i,j,l} {{e^{i\left( {{\bf{k}}{{\bf{r}}_{jl}} + {{\bf{k}}^\prime }{{\bf{r}}_{li}}} \right)}}} H_0^{(1)}\left( {k{r_{ij}}} \right)\left[ {{\delta _{\alpha \beta }}\chi _z^{lij} + \left( {1 - {\delta _{\alpha \beta }}} \right)\left( {\chi _x^{lij} + \chi _y^{lij}} \right)} \right],
%	&{\left| {{T_{{\bf k'k}\alpha \beta }}} \right|^2}_{chiral} = \frac{{{A^3}2{\pi %^2}m}}{{{\hbar ^2}}}{\rm{Re}}\sum\limits_{i,j,l} {{e^{i\left( {{\bf{k}}{{\bf{r}}_{jl}} %+ {{\bf{k}}^\prime }{{\bf{r}}_{li}}} \right)}}} H_0^{(1)}\left( {k{r_{ij}}} \right)%
%	\left( \delta_{\alpha \beta} \chi_z^{lij} + (1-\delta_{\alpha \beta}) %\left[\chi_x^{lij} +  \chi_y^{lij} \right]\right)
	%
	\end{align}
\end{widetext}
where $\delta_{\alpha\beta}$ is the Kronecker delta.
The spin chirality driven term does not depend on the %incident electron spin:
%We emphasize that the expression (\ref{eqTchiral}) does not depend on the
incident electron spin: ${\left| {{T_{ \uparrow  \uparrow }}} \right|^2_{\chi}} = {\left| {{T_{ \downarrow  \downarrow }}} \right|^2_{\chi}}$, ${\left| {{T_{ \uparrow  \downarrow }}} \right|^2_{\chi}} = {\left| {{T_{ \downarrow  \uparrow }}} \right|^2_{\chi}}$.
%It is now obvious that sign of the spin-chirality driven term does not depend on the %incident electron spin: ${\left| {{T_{ \uparrow  \uparrow }}} \right|^2} = {\left| %{{T_{ \downarrow  \downarrow }}} \right|^2}$, ${\left| {{T_{ \uparrow  \downarrow }}} %\right|^2} = {\left| {{T_{ \downarrow  \uparrow }}} \right|^2}$.
Let us emphasize the origin of this result. The spin chirality contribution is always due to the interference between one spin-conserving and two spin-flip scattering events.
While for the spin-conserving scattering amplitude the sign is opposite for spin-up and spin-down states (\ref{eqvij}), it is compensated by the sign change for the double spin flip process.
%
%in Eq.() due to different bypass order in double spin-flip process.

\bibliography{Skyrmion}

\begin{thebibliography}{54}
\expandafter\ifx\csname natexlab\endcsname\relax\def\natexlab#1{#1}\fi
\expandafter\ifx\csname bibnamefont\endcsname\relax
  \def\bibnamefont#1{#1}\fi
\expandafter\ifx\csname bibfnamefont\endcsname\relax
  \def\bibfnamefont#1{#1}\fi
\expandafter\ifx\csname citenamefont\endcsname\relax
  \def\citenamefont#1{#1}\fi
\expandafter\ifx\csname url\endcsname\relax
  \def\url#1{\texttt{#1}}\fi
\expandafter\ifx\csname urlprefix\endcsname\relax\def\urlprefix{URL }\fi
\providecommand{\bibinfo}[2]{#2}
\providecommand{\eprint}[2][]{\url{#2}}

\bibitem[{\citenamefont{Dyakonov}(2008)}]{Dyakonov}
\bibinfo{author}{\bibfnamefont{M.}~\bibnamefont{Dyakonov}}, in
  \emph{\bibinfo{booktitle}{Spin Physics in Semiconductors}}
  (\bibinfo{publisher}{Springer}, \bibinfo{year}{2008}).

\bibitem[{\citenamefont{Nagaosa et~al.}(2010)\citenamefont{Nagaosa, Sinova,
  Onoda, MacDonald, and Ong}}]{AHE-Sinova}
\bibinfo{author}{\bibfnamefont{N.}~\bibnamefont{Nagaosa}},
  \bibinfo{author}{\bibfnamefont{J.}~\bibnamefont{Sinova}},
  \bibinfo{author}{\bibfnamefont{S.}~\bibnamefont{Onoda}},
  \bibinfo{author}{\bibfnamefont{A.~H.} \bibnamefont{MacDonald}},
  \bibnamefont{and} \bibinfo{author}{\bibfnamefont{N.~P.} \bibnamefont{Ong}},
  \bibinfo{journal}{Rev. Mod. Phys.} \textbf{\bibinfo{volume}{82}},
  \bibinfo{pages}{1539} (\bibinfo{year}{2010}).

\bibitem[{\citenamefont{Sinitsyn}(2008)}]{Sinitsyn}
\bibinfo{author}{\bibfnamefont{N.~A.} \bibnamefont{Sinitsyn}},
  \bibinfo{journal}{Journal of Physics: Condensed Matter}
  \textbf{\bibinfo{volume}{20}}, \bibinfo{pages}{023201}
  (\bibinfo{year}{2008}).

\bibitem[{\citenamefont{Abakumov and Yassievich}(1972)}]{AbakumovAHE}
\bibinfo{author}{\bibfnamefont{V.}~\bibnamefont{Abakumov}} \bibnamefont{and}
  \bibinfo{author}{\bibfnamefont{I.}~\bibnamefont{Yassievich}},
  \bibinfo{journal}{Soviet Physics JETP} \textbf{\bibinfo{volume}{34}},
  \bibinfo{pages}{1375} (\bibinfo{year}{1972}).

\bibitem[{\citenamefont{Neubauer et~al.}(2009)\citenamefont{Neubauer,
  Pfleiderer, Binz, Rosch, Ritz, Niklowitz, and B\"oni}}]{MnSiAPhase}
\bibinfo{author}{\bibfnamefont{A.}~\bibnamefont{Neubauer}},
  \bibinfo{author}{\bibfnamefont{C.}~\bibnamefont{Pfleiderer}},
  \bibinfo{author}{\bibfnamefont{B.}~\bibnamefont{Binz}},
  \bibinfo{author}{\bibfnamefont{A.}~\bibnamefont{Rosch}},
  \bibinfo{author}{\bibfnamefont{R.}~\bibnamefont{Ritz}},
  \bibinfo{author}{\bibfnamefont{P.~G.} \bibnamefont{Niklowitz}},
  \bibnamefont{and} \bibinfo{author}{\bibfnamefont{P.}~\bibnamefont{B\"oni}},
  \bibinfo{journal}{Phys. Rev. Lett.} \textbf{\bibinfo{volume}{102}},
  \bibinfo{pages}{186602} (\bibinfo{year}{2009}).

\bibitem[{\citenamefont{Lee et~al.}(2009)\citenamefont{Lee, Kang, Onose,
  Tokura, and Ong}}]{LeeOnose}
\bibinfo{author}{\bibfnamefont{M.}~\bibnamefont{Lee}},
  \bibinfo{author}{\bibfnamefont{W.}~\bibnamefont{Kang}},
  \bibinfo{author}{\bibfnamefont{Y.}~\bibnamefont{Onose}},
  \bibinfo{author}{\bibfnamefont{Y.}~\bibnamefont{Tokura}}, \bibnamefont{and}
  \bibinfo{author}{\bibfnamefont{N.~P.} \bibnamefont{Ong}},
  \bibinfo{journal}{Phys. Rev. Lett.} \textbf{\bibinfo{volume}{102}},
  \bibinfo{pages}{186601} (\bibinfo{year}{2009}).

\bibitem[{\citenamefont{Lobanov et~al.}(2016)\citenamefont{Lobanov,
  J{\'o}nsson, and Uzdin}}]{Uzdin-1}
\bibinfo{author}{\bibfnamefont{I.~S.} \bibnamefont{Lobanov}},
  \bibinfo{author}{\bibfnamefont{H.}~\bibnamefont{J{\'o}nsson}},
  \bibnamefont{and} \bibinfo{author}{\bibfnamefont{V.~M.} \bibnamefont{Uzdin}},
  \bibinfo{journal}{Physical Review B} \textbf{\bibinfo{volume}{94}},
  \bibinfo{pages}{174418} (\bibinfo{year}{2016}).

\bibitem[{\citenamefont{Hagemeister et~al.}(2015)\citenamefont{Hagemeister,
  Romming, Von~Bergmann, Vedmedenko, and Wiesendanger}}]{Wiesendanger-1}
\bibinfo{author}{\bibfnamefont{J.}~\bibnamefont{Hagemeister}},
  \bibinfo{author}{\bibfnamefont{N.}~\bibnamefont{Romming}},
  \bibinfo{author}{\bibfnamefont{K.}~\bibnamefont{Von~Bergmann}},
  \bibinfo{author}{\bibfnamefont{E.}~\bibnamefont{Vedmedenko}},
  \bibnamefont{and}
  \bibinfo{author}{\bibfnamefont{R.}~\bibnamefont{Wiesendanger}},
  \bibinfo{journal}{Nature communications} \textbf{\bibinfo{volume}{6}}
  (\bibinfo{year}{2015}).

\bibitem[{\citenamefont{R{\'o}zsa et~al.}(2016)\citenamefont{R{\'o}zsa, Simon,
  Palot{\'a}s, Udvardi, and Szunyogh}}]{rozsa2016complex}
\bibinfo{author}{\bibfnamefont{L.}~\bibnamefont{R{\'o}zsa}},
  \bibinfo{author}{\bibfnamefont{E.}~\bibnamefont{Simon}},
  \bibinfo{author}{\bibfnamefont{K.}~\bibnamefont{Palot{\'a}s}},
  \bibinfo{author}{\bibfnamefont{L.}~\bibnamefont{Udvardi}}, \bibnamefont{and}
  \bibinfo{author}{\bibfnamefont{L.}~\bibnamefont{Szunyogh}},
  \bibinfo{journal}{Physical Review B} \textbf{\bibinfo{volume}{93}},
  \bibinfo{pages}{024417} (\bibinfo{year}{2016}).

\bibitem[{\citenamefont{Taguchi et~al.}(2001)\citenamefont{Taguchi, Oohara,
  Yoshizawa, Nagaosa, and Tokura}}]{Taguchi_Science}
\bibinfo{author}{\bibfnamefont{Y.}~\bibnamefont{Taguchi}},
  \bibinfo{author}{\bibfnamefont{Y.}~\bibnamefont{Oohara}},
  \bibinfo{author}{\bibfnamefont{H.}~\bibnamefont{Yoshizawa}},
  \bibinfo{author}{\bibfnamefont{N.}~\bibnamefont{Nagaosa}}, \bibnamefont{and}
  \bibinfo{author}{\bibfnamefont{Y.}~\bibnamefont{Tokura}},
  \bibinfo{journal}{Science} \textbf{\bibinfo{volume}{291}},
  \bibinfo{pages}{2573} (\bibinfo{year}{2001}).

\bibitem[{\citenamefont{Machida et~al.}(2007)\citenamefont{Machida, Nakatsuji,
  Maeno, Tayama, Sakakibara, and Onoda}}]{Machida_PRL}
\bibinfo{author}{\bibfnamefont{Y.}~\bibnamefont{Machida}},
  \bibinfo{author}{\bibfnamefont{S.}~\bibnamefont{Nakatsuji}},
  \bibinfo{author}{\bibfnamefont{Y.}~\bibnamefont{Maeno}},
  \bibinfo{author}{\bibfnamefont{T.}~\bibnamefont{Tayama}},
  \bibinfo{author}{\bibfnamefont{T.}~\bibnamefont{Sakakibara}},
  \bibnamefont{and} \bibinfo{author}{\bibfnamefont{S.}~\bibnamefont{Onoda}},
  \bibinfo{journal}{Phys. Rev. Lett.} \textbf{\bibinfo{volume}{98}},
  \bibinfo{pages}{057203} (\bibinfo{year}{2007}).

\bibitem[{\citenamefont{S{\"u}rgers et~al.}(2014)\citenamefont{S{\"u}rgers,
  Fischer, Winkel, and L{\"o}hneysen}}]{surgers2014large}
\bibinfo{author}{\bibfnamefont{C.}~\bibnamefont{S{\"u}rgers}},
  \bibinfo{author}{\bibfnamefont{G.}~\bibnamefont{Fischer}},
  \bibinfo{author}{\bibfnamefont{P.}~\bibnamefont{Winkel}}, \bibnamefont{and}
  \bibinfo{author}{\bibfnamefont{H.~v.} \bibnamefont{L{\"o}hneysen}},
  \bibinfo{journal}{Nature communications} \textbf{\bibinfo{volume}{5}}
  (\bibinfo{year}{2014}).

\bibitem[{\citenamefont{Ueland et~al.}(2012)\citenamefont{Ueland, Miclea, Kato,
  Ayala-Valenzuela, McDonald, Okazaki, Tobash, Torrez, Ronning, Movshovich
  et~al.}}]{ueland2012controllable}
\bibinfo{author}{\bibfnamefont{B.}~\bibnamefont{Ueland}},
  \bibinfo{author}{\bibfnamefont{C.}~\bibnamefont{Miclea}},
  \bibinfo{author}{\bibfnamefont{Y.}~\bibnamefont{Kato}},
  \bibinfo{author}{\bibfnamefont{O.}~\bibnamefont{Ayala-Valenzuela}},
  \bibinfo{author}{\bibfnamefont{R.}~\bibnamefont{McDonald}},
  \bibinfo{author}{\bibfnamefont{R.}~\bibnamefont{Okazaki}},
  \bibinfo{author}{\bibfnamefont{P.}~\bibnamefont{Tobash}},
  \bibinfo{author}{\bibfnamefont{M.}~\bibnamefont{Torrez}},
  \bibinfo{author}{\bibfnamefont{F.}~\bibnamefont{Ronning}},
  \bibinfo{author}{\bibfnamefont{R.}~\bibnamefont{Movshovich}},
  \bibnamefont{et~al.}, \bibinfo{journal}{Nature communications}
  \textbf{\bibinfo{volume}{3}}, \bibinfo{pages}{1067} (\bibinfo{year}{2012}).

\bibitem[{\citenamefont{Fabris et~al.}(2006)\citenamefont{Fabris, Pureur,
  Schaf, Vieira, and Campbell}}]{SpinGlass1}
\bibinfo{author}{\bibfnamefont{F.~W.} \bibnamefont{Fabris}},
  \bibinfo{author}{\bibfnamefont{P.}~\bibnamefont{Pureur}},
  \bibinfo{author}{\bibfnamefont{J.}~\bibnamefont{Schaf}},
  \bibinfo{author}{\bibfnamefont{V.~N.} \bibnamefont{Vieira}},
  \bibnamefont{and} \bibinfo{author}{\bibfnamefont{I.~A.}
  \bibnamefont{Campbell}}, \bibinfo{journal}{Phys. Rev. B}
  \textbf{\bibinfo{volume}{74}}, \bibinfo{pages}{214201}
  (\bibinfo{year}{2006}).

\bibitem[{\citenamefont{Taniguchi et~al.}(2004)\citenamefont{Taniguchi,
  Yamanaka, Sumioka, Yamazaki, Tabata, and Kawarazaki}}]{SpinGlass2}
\bibinfo{author}{\bibfnamefont{T.}~\bibnamefont{Taniguchi}},
  \bibinfo{author}{\bibfnamefont{K.}~\bibnamefont{Yamanaka}},
  \bibinfo{author}{\bibfnamefont{H.}~\bibnamefont{Sumioka}},
  \bibinfo{author}{\bibfnamefont{T.}~\bibnamefont{Yamazaki}},
  \bibinfo{author}{\bibfnamefont{Y.}~\bibnamefont{Tabata}}, \bibnamefont{and}
  \bibinfo{author}{\bibfnamefont{S.}~\bibnamefont{Kawarazaki}},
  \bibinfo{journal}{Phys. Rev. Lett.} \textbf{\bibinfo{volume}{93}},
  \bibinfo{pages}{246605} (\bibinfo{year}{2004}).

\bibitem[{\citenamefont{Ohuchi et~al.}(2015)\citenamefont{Ohuchi, Kozuka,
  Uchida, Ueno, Tsukazaki, and Kawasaki}}]{EuO}
\bibinfo{author}{\bibfnamefont{Y.}~\bibnamefont{Ohuchi}},
  \bibinfo{author}{\bibfnamefont{Y.}~\bibnamefont{Kozuka}},
  \bibinfo{author}{\bibfnamefont{M.}~\bibnamefont{Uchida}},
  \bibinfo{author}{\bibfnamefont{K.}~\bibnamefont{Ueno}},
  \bibinfo{author}{\bibfnamefont{A.}~\bibnamefont{Tsukazaki}},
  \bibnamefont{and} \bibinfo{author}{\bibfnamefont{M.}~\bibnamefont{Kawasaki}},
  \bibinfo{journal}{Phys. Rev. B} \textbf{\bibinfo{volume}{91}},
  \bibinfo{pages}{245115} (\bibinfo{year}{2015}).

\bibitem[{\citenamefont{Matl et~al.}(1998)\citenamefont{Matl, Ong, Yan, Li,
  Studebaker, Baum, and Doubinina}}]{Matl_CMR}
\bibinfo{author}{\bibfnamefont{P.}~\bibnamefont{Matl}},
  \bibinfo{author}{\bibfnamefont{N.~P.} \bibnamefont{Ong}},
  \bibinfo{author}{\bibfnamefont{Y.~F.} \bibnamefont{Yan}},
  \bibinfo{author}{\bibfnamefont{Y.~Q.} \bibnamefont{Li}},
  \bibinfo{author}{\bibfnamefont{D.}~\bibnamefont{Studebaker}},
  \bibinfo{author}{\bibfnamefont{T.}~\bibnamefont{Baum}}, \bibnamefont{and}
  \bibinfo{author}{\bibfnamefont{G.}~\bibnamefont{Doubinina}},
  \bibinfo{journal}{Phys. Rev. B} \textbf{\bibinfo{volume}{57}},
  \bibinfo{pages}{10248} (\bibinfo{year}{1998}).

\bibitem[{\citenamefont{Jakob et~al.}(1998)\citenamefont{Jakob, Martin,
  Westerburg, and Adrian}}]{Jacob_CMR}
\bibinfo{author}{\bibfnamefont{G.}~\bibnamefont{Jakob}},
  \bibinfo{author}{\bibfnamefont{F.}~\bibnamefont{Martin}},
  \bibinfo{author}{\bibfnamefont{W.}~\bibnamefont{Westerburg}},
  \bibnamefont{and} \bibinfo{author}{\bibfnamefont{H.}~\bibnamefont{Adrian}},
  \bibinfo{journal}{Phys. Rev. B} \textbf{\bibinfo{volume}{57}},
  \bibinfo{pages}{10252} (\bibinfo{year}{1998}).

\bibitem[{\citenamefont{Oveshnikov et~al.}(2015)\citenamefont{Oveshnikov,
  Kulbachinskii, Davydov, Aronzon, Rozhansky, Averkiev, Kugel, and
  Tripathi}}]{AronzonRozh}
\bibinfo{author}{\bibfnamefont{L.~N.} \bibnamefont{Oveshnikov}},
  \bibinfo{author}{\bibfnamefont{V.~A.} \bibnamefont{Kulbachinskii}},
  \bibinfo{author}{\bibfnamefont{A.~B.} \bibnamefont{Davydov}},
  \bibinfo{author}{\bibfnamefont{B.~A.} \bibnamefont{Aronzon}},
  \bibinfo{author}{\bibfnamefont{I.~V.} \bibnamefont{Rozhansky}},
  \bibinfo{author}{\bibfnamefont{N.~S.} \bibnamefont{Averkiev}},
  \bibinfo{author}{\bibfnamefont{K.~I.} \bibnamefont{Kugel}}, \bibnamefont{and}
  \bibinfo{author}{\bibfnamefont{V.}~\bibnamefont{Tripathi}},
  \bibinfo{journal}{Scientific Reports} \textbf{\bibinfo{volume}{5}},
  \bibinfo{pages}{17158} (\bibinfo{year}{2015}).

\bibitem[{\citenamefont{Nagaosa and Tokura}(2013)}]{NagaosaNature}
\bibinfo{author}{\bibfnamefont{N.}~\bibnamefont{Nagaosa}} \bibnamefont{and}
  \bibinfo{author}{\bibfnamefont{Y.}~\bibnamefont{Tokura}},
  \bibinfo{journal}{Nature Nanotechnoloy} \textbf{\bibinfo{volume}{8}},
  \bibinfo{pages}{899} (\bibinfo{year}{2013}).

\bibitem[{\citenamefont{Kanazawa et~al.}(2015)\citenamefont{Kanazawa, Kubota,
  Tsukazaki, Kozuka, Takahashi, Kawasaki, Ichikawa, Kagawa, and
  Tokura}}]{DiscretHall}
\bibinfo{author}{\bibfnamefont{N.}~\bibnamefont{Kanazawa}},
  \bibinfo{author}{\bibfnamefont{M.}~\bibnamefont{Kubota}},
  \bibinfo{author}{\bibfnamefont{A.}~\bibnamefont{Tsukazaki}},
  \bibinfo{author}{\bibfnamefont{Y.}~\bibnamefont{Kozuka}},
  \bibinfo{author}{\bibfnamefont{K.~S.} \bibnamefont{Takahashi}},
  \bibinfo{author}{\bibfnamefont{M.}~\bibnamefont{Kawasaki}},
  \bibinfo{author}{\bibfnamefont{M.}~\bibnamefont{Ichikawa}},
  \bibinfo{author}{\bibfnamefont{F.}~\bibnamefont{Kagawa}}, \bibnamefont{and}
  \bibinfo{author}{\bibfnamefont{Y.}~\bibnamefont{Tokura}},
  \bibinfo{journal}{Phys. Rev. B} \textbf{\bibinfo{volume}{91}},
  \bibinfo{pages}{041122} (\bibinfo{year}{2015}).

\bibitem[{\citenamefont{Mühlbauer et~al.}(2009)\citenamefont{Mühlbauer, Binz,
  Jonietz, Pfleiderer, Rosch, Neubauer, Georgii, and
  Böni}}]{Muhl_MnSi_Science}
\bibinfo{author}{\bibfnamefont{S.}~\bibnamefont{Mühlbauer}},
  \bibinfo{author}{\bibfnamefont{B.}~\bibnamefont{Binz}},
  \bibinfo{author}{\bibfnamefont{F.}~\bibnamefont{Jonietz}},
  \bibinfo{author}{\bibfnamefont{C.}~\bibnamefont{Pfleiderer}},
  \bibinfo{author}{\bibfnamefont{A.}~\bibnamefont{Rosch}},
  \bibinfo{author}{\bibfnamefont{A.}~\bibnamefont{Neubauer}},
  \bibinfo{author}{\bibfnamefont{R.}~\bibnamefont{Georgii}}, \bibnamefont{and}
  \bibinfo{author}{\bibfnamefont{P.}~\bibnamefont{Böni}},
  \bibinfo{journal}{Science} \textbf{\bibinfo{volume}{323}},
  \bibinfo{pages}{915} (\bibinfo{year}{2009}).

\bibitem[{\citenamefont{Li et~al.}(2013)\citenamefont{Li, Kanazawa, Yu,
  Tsukazaki, Kawasaki, Ichikawa, Jin, Kagawa, and Tokura}}]{THE_Li}
\bibinfo{author}{\bibfnamefont{Y.}~\bibnamefont{Li}},
  \bibinfo{author}{\bibfnamefont{N.}~\bibnamefont{Kanazawa}},
  \bibinfo{author}{\bibfnamefont{X.~Z.} \bibnamefont{Yu}},
  \bibinfo{author}{\bibfnamefont{A.}~\bibnamefont{Tsukazaki}},
  \bibinfo{author}{\bibfnamefont{M.}~\bibnamefont{Kawasaki}},
  \bibinfo{author}{\bibfnamefont{M.}~\bibnamefont{Ichikawa}},
  \bibinfo{author}{\bibfnamefont{X.~F.} \bibnamefont{Jin}},
  \bibinfo{author}{\bibfnamefont{F.}~\bibnamefont{Kagawa}}, \bibnamefont{and}
  \bibinfo{author}{\bibfnamefont{Y.}~\bibnamefont{Tokura}},
  \bibinfo{journal}{Phys. Rev. Lett.} \textbf{\bibinfo{volume}{110}},
  \bibinfo{pages}{117202} (\bibinfo{year}{2013}).

\bibitem[{\citenamefont{Chapman et~al.}(2013)\citenamefont{Chapman,
  Grossnickle, Wolf, and Lee}}]{Chapman_PRB}
\bibinfo{author}{\bibfnamefont{B.~J.} \bibnamefont{Chapman}},
  \bibinfo{author}{\bibfnamefont{M.~G.} \bibnamefont{Grossnickle}},
  \bibinfo{author}{\bibfnamefont{T.}~\bibnamefont{Wolf}}, \bibnamefont{and}
  \bibinfo{author}{\bibfnamefont{M.}~\bibnamefont{Lee}},
  \bibinfo{journal}{Phys. Rev. B} \textbf{\bibinfo{volume}{88}},
  \bibinfo{pages}{214406} (\bibinfo{year}{2013}).

\bibitem[{\citenamefont{M\"unzer et~al.}(2010)\citenamefont{M\"unzer, Neubauer,
  Adams, M\"uhlbauer, Franz, Jonietz, Georgii, B\"oni, Pedersen, Schmidt
  et~al.}}]{Munzer_PRB}
\bibinfo{author}{\bibfnamefont{W.}~\bibnamefont{M\"unzer}},
  \bibinfo{author}{\bibfnamefont{A.}~\bibnamefont{Neubauer}},
  \bibinfo{author}{\bibfnamefont{T.}~\bibnamefont{Adams}},
  \bibinfo{author}{\bibfnamefont{S.}~\bibnamefont{M\"uhlbauer}},
  \bibinfo{author}{\bibfnamefont{C.}~\bibnamefont{Franz}},
  \bibinfo{author}{\bibfnamefont{F.}~\bibnamefont{Jonietz}},
  \bibinfo{author}{\bibfnamefont{R.}~\bibnamefont{Georgii}},
  \bibinfo{author}{\bibfnamefont{P.}~\bibnamefont{B\"oni}},
  \bibinfo{author}{\bibfnamefont{B.}~\bibnamefont{Pedersen}},
  \bibinfo{author}{\bibfnamefont{M.}~\bibnamefont{Schmidt}},
  \bibnamefont{et~al.}, \bibinfo{journal}{Phys. Rev. B}
  \textbf{\bibinfo{volume}{81}}, \bibinfo{pages}{041203}
  (\bibinfo{year}{2010}).

\bibitem[{\citenamefont{Yu et~al.}(2011)\citenamefont{Yu, Kanazawa, Onose,
  Kimoto, Zhang, Ishiwata, Matsui, and Tokura}}]{FeGe_THE}
\bibinfo{author}{\bibfnamefont{X.~Z.} \bibnamefont{Yu}},
  \bibinfo{author}{\bibfnamefont{N.}~\bibnamefont{Kanazawa}},
  \bibinfo{author}{\bibfnamefont{Y.}~\bibnamefont{Onose}},
  \bibinfo{author}{\bibfnamefont{K.}~\bibnamefont{Kimoto}},
  \bibinfo{author}{\bibfnamefont{W.~Z.} \bibnamefont{Zhang}},
  \bibinfo{author}{\bibfnamefont{S.}~\bibnamefont{Ishiwata}},
  \bibinfo{author}{\bibfnamefont{Y.}~\bibnamefont{Matsui}}, \bibnamefont{and}
  \bibinfo{author}{\bibfnamefont{Y.}~\bibnamefont{Tokura}},
  \bibinfo{journal}{Nature Materials} \textbf{\bibinfo{volume}{10}},
  \bibinfo{pages}{106} (\bibinfo{year}{2011}).

\bibitem[{\citenamefont{Zhou and Ezawa}(2014)}]{Zhou-1}
\bibinfo{author}{\bibfnamefont{Y.}~\bibnamefont{Zhou}} \bibnamefont{and}
  \bibinfo{author}{\bibfnamefont{M.}~\bibnamefont{Ezawa}},
  \bibinfo{journal}{Nat Commun} \textbf{\bibinfo{volume}{5}},
  \bibinfo{pages}{4652} (\bibinfo{year}{2014}).

\bibitem[{\citenamefont{Ma et~al.}(2015)\citenamefont{Ma, Zhou, Braun, and
  Lew}}]{Zhou-2}
\bibinfo{author}{\bibfnamefont{F.}~\bibnamefont{Ma}},
  \bibinfo{author}{\bibfnamefont{Y.}~\bibnamefont{Zhou}},
  \bibinfo{author}{\bibfnamefont{H.~B.} \bibnamefont{Braun}}, \bibnamefont{and}
  \bibinfo{author}{\bibfnamefont{W.~S.} \bibnamefont{Lew}},
  \bibinfo{journal}{Nano Letters} \textbf{\bibinfo{volume}{15}},
  \bibinfo{pages}{4029} (\bibinfo{year}{2015}).

\bibitem[{\citenamefont{Gilbert et~al.}(2015)\citenamefont{Gilbert, Maranville,
  Balk, Kirby, Fischer, Pierce, Unguris, Borchers, and
  Liu}}]{gilbert2015realization}
\bibinfo{author}{\bibfnamefont{D.~A.} \bibnamefont{Gilbert}},
  \bibinfo{author}{\bibfnamefont{B.~B.} \bibnamefont{Maranville}},
  \bibinfo{author}{\bibfnamefont{A.~L.} \bibnamefont{Balk}},
  \bibinfo{author}{\bibfnamefont{B.~J.} \bibnamefont{Kirby}},
  \bibinfo{author}{\bibfnamefont{P.}~\bibnamefont{Fischer}},
  \bibinfo{author}{\bibfnamefont{D.~T.} \bibnamefont{Pierce}},
  \bibinfo{author}{\bibfnamefont{J.}~\bibnamefont{Unguris}},
  \bibinfo{author}{\bibfnamefont{J.~A.} \bibnamefont{Borchers}},
  \bibnamefont{and} \bibinfo{author}{\bibfnamefont{K.}~\bibnamefont{Liu}},
  \bibinfo{journal}{Nature communications} \textbf{\bibinfo{volume}{6}}
  (\bibinfo{year}{2015}).

\bibitem[{\citenamefont{Li et~al.}(2014)\citenamefont{Li, Tan, Moon, Doran,
  Marcus, Young, Arenholz, Ma, Yang, Hwang et~al.}}]{li2014tailoring}
\bibinfo{author}{\bibfnamefont{J.}~\bibnamefont{Li}},
  \bibinfo{author}{\bibfnamefont{A.}~\bibnamefont{Tan}},
  \bibinfo{author}{\bibfnamefont{K.}~\bibnamefont{Moon}},
  \bibinfo{author}{\bibfnamefont{A.}~\bibnamefont{Doran}},
  \bibinfo{author}{\bibfnamefont{M.}~\bibnamefont{Marcus}},
  \bibinfo{author}{\bibfnamefont{A.}~\bibnamefont{Young}},
  \bibinfo{author}{\bibfnamefont{E.}~\bibnamefont{Arenholz}},
  \bibinfo{author}{\bibfnamefont{S.}~\bibnamefont{Ma}},
  \bibinfo{author}{\bibfnamefont{R.}~\bibnamefont{Yang}},
  \bibinfo{author}{\bibfnamefont{C.}~\bibnamefont{Hwang}},
  \bibnamefont{et~al.}, \bibinfo{journal}{Nature communications}
  \textbf{\bibinfo{volume}{5}} (\bibinfo{year}{2014}).

\bibitem[{\citenamefont{{Nayak} et~al.}(2017)\citenamefont{{Nayak}, {Kumar},
  {Werner}, {Pippel}, {Sahoo}, {Damay}, {R{\"o}{\ss}ler}, {Felser}, and
  {Parkin}}}]{Parkin}
\bibinfo{author}{\bibfnamefont{A.~K.} \bibnamefont{{Nayak}}},
  \bibinfo{author}{\bibfnamefont{V.}~\bibnamefont{{Kumar}}},
  \bibinfo{author}{\bibfnamefont{P.}~\bibnamefont{{Werner}}},
  \bibinfo{author}{\bibfnamefont{E.}~\bibnamefont{{Pippel}}},
  \bibinfo{author}{\bibfnamefont{R.}~\bibnamefont{{Sahoo}}},
  \bibinfo{author}{\bibfnamefont{F.}~\bibnamefont{{Damay}}},
  \bibinfo{author}{\bibfnamefont{U.~K.} \bibnamefont{{R{\"o}{\ss}ler}}},
  \bibinfo{author}{\bibfnamefont{C.}~\bibnamefont{{Felser}}}, \bibnamefont{and}
  \bibinfo{author}{\bibfnamefont{S.~S.~P.} \bibnamefont{{Parkin}}},
  \bibinfo{journal}{ArXiv e-prints}  (\bibinfo{year}{2017}),
  \eprint{1703.01017}.

\bibitem[{\citenamefont{Kang et~al.}(2016)\citenamefont{Kang, Huang, Zheng, Lv,
  Lei, Zhang, Zhang, Zhou, and Zhao}}]{racetrack}
\bibinfo{author}{\bibfnamefont{W.}~\bibnamefont{Kang}},
  \bibinfo{author}{\bibfnamefont{Y.}~\bibnamefont{Huang}},
  \bibinfo{author}{\bibfnamefont{C.}~\bibnamefont{Zheng}},
  \bibinfo{author}{\bibfnamefont{W.}~\bibnamefont{Lv}},
  \bibinfo{author}{\bibfnamefont{N.}~\bibnamefont{Lei}},
  \bibinfo{author}{\bibfnamefont{Y.}~\bibnamefont{Zhang}},
  \bibinfo{author}{\bibfnamefont{X.}~\bibnamefont{Zhang}},
  \bibinfo{author}{\bibfnamefont{Y.}~\bibnamefont{Zhou}}, \bibnamefont{and}
  \bibinfo{author}{\bibfnamefont{W.}~\bibnamefont{Zhao}},
  \bibinfo{journal}{Scientific reports} \textbf{\bibinfo{volume}{6}}
  (\bibinfo{year}{2016}).

\bibitem[{\citenamefont{Liang et~al.}(2015)\citenamefont{Liang, DeGrave, Stolt,
  Tokura, and Jin}}]{liang2015current}
\bibinfo{author}{\bibfnamefont{D.}~\bibnamefont{Liang}},
  \bibinfo{author}{\bibfnamefont{J.~P.} \bibnamefont{DeGrave}},
  \bibinfo{author}{\bibfnamefont{M.~J.} \bibnamefont{Stolt}},
  \bibinfo{author}{\bibfnamefont{Y.}~\bibnamefont{Tokura}}, \bibnamefont{and}
  \bibinfo{author}{\bibfnamefont{S.}~\bibnamefont{Jin}},
  \bibinfo{journal}{Nature communications} \textbf{\bibinfo{volume}{6}}
  (\bibinfo{year}{2015}).

\bibitem[{\citenamefont{Tomasello et~al.}(2014)\citenamefont{Tomasello,
  Martinez, Zivieri, Torres, Carpentieri, and
  Finocchio}}]{tomasello2014strategy}
\bibinfo{author}{\bibfnamefont{R.}~\bibnamefont{Tomasello}},
  \bibinfo{author}{\bibfnamefont{E.}~\bibnamefont{Martinez}},
  \bibinfo{author}{\bibfnamefont{R.}~\bibnamefont{Zivieri}},
  \bibinfo{author}{\bibfnamefont{L.}~\bibnamefont{Torres}},
  \bibinfo{author}{\bibfnamefont{M.}~\bibnamefont{Carpentieri}},
  \bibnamefont{and}
  \bibinfo{author}{\bibfnamefont{G.}~\bibnamefont{Finocchio}},
  \bibinfo{journal}{Scientific Reports} \textbf{\bibinfo{volume}{4}},
  \bibinfo{pages}{6784} (\bibinfo{year}{2014}).

\bibitem[{\citenamefont{{Maccariello} et~al.}(2017)\citenamefont{{Maccariello},
  {Legrand}, {Reyren}, {Garcia}, {Bouzehouane}, {Collin}, {Cros}, and
  {Fert}}}]{Fert}
\bibinfo{author}{\bibfnamefont{D.}~\bibnamefont{{Maccariello}}},
  \bibinfo{author}{\bibfnamefont{W.}~\bibnamefont{{Legrand}}},
  \bibinfo{author}{\bibfnamefont{N.}~\bibnamefont{{Reyren}}},
  \bibinfo{author}{\bibfnamefont{K.}~\bibnamefont{{Garcia}}},
  \bibinfo{author}{\bibfnamefont{K.}~\bibnamefont{{Bouzehouane}}},
  \bibinfo{author}{\bibfnamefont{S.}~\bibnamefont{{Collin}}},
  \bibinfo{author}{\bibfnamefont{V.}~\bibnamefont{{Cros}}}, \bibnamefont{and}
  \bibinfo{author}{\bibfnamefont{A.}~\bibnamefont{{Fert}}},
  \bibinfo{journal}{ArXiv e-prints}  (\bibinfo{year}{2017}),
  \eprint{1706.05809}.

\bibitem[{\citenamefont{Zhang et~al.}(2015)\citenamefont{Zhang, Zhao, Fangohr,
  Liu, Xia, Xia, and Morvan}}]{sk_memory}
\bibinfo{author}{\bibfnamefont{X.}~\bibnamefont{Zhang}},
  \bibinfo{author}{\bibfnamefont{G.}~\bibnamefont{Zhao}},
  \bibinfo{author}{\bibfnamefont{H.}~\bibnamefont{Fangohr}},
  \bibinfo{author}{\bibfnamefont{J.~P.} \bibnamefont{Liu}},
  \bibinfo{author}{\bibfnamefont{W.}~\bibnamefont{Xia}},
  \bibinfo{author}{\bibfnamefont{J.}~\bibnamefont{Xia}}, \bibnamefont{and}
  \bibinfo{author}{\bibfnamefont{F.}~\bibnamefont{Morvan}},
  \bibinfo{journal}{Scientific Reports} \textbf{\bibinfo{volume}{5}},
  \bibinfo{pages}{7643} (\bibinfo{year}{2015}).

\bibitem[{\citenamefont{Fert et~al.}(2013)\citenamefont{Fert, Cros, and
  Sampaio}}]{fert2013skyrmions}
\bibinfo{author}{\bibfnamefont{A.}~\bibnamefont{Fert}},
  \bibinfo{author}{\bibfnamefont{V.}~\bibnamefont{Cros}}, \bibnamefont{and}
  \bibinfo{author}{\bibfnamefont{J.}~\bibnamefont{Sampaio}},
  \bibinfo{journal}{Nature nanotechnology} \textbf{\bibinfo{volume}{8}},
  \bibinfo{pages}{152} (\bibinfo{year}{2013}).

\bibitem[{\citenamefont{Zhang et~al.}(2014)\citenamefont{Zhang, Ezawa, and
  Zhou}}]{zhang2014magnetic}
\bibinfo{author}{\bibfnamefont{X.}~\bibnamefont{Zhang}},
  \bibinfo{author}{\bibfnamefont{M.}~\bibnamefont{Ezawa}}, \bibnamefont{and}
  \bibinfo{author}{\bibfnamefont{Y.}~\bibnamefont{Zhou}},
  \bibinfo{journal}{Scientific Reports} \textbf{\bibinfo{volume}{5}},
  \bibinfo{pages}{9400} (\bibinfo{year}{2014}).

\bibitem[{\citenamefont{Bruno et~al.}(2004)\citenamefont{Bruno, Dugaev, and
  Taillefumier}}]{BrunoDugaev}
\bibinfo{author}{\bibfnamefont{P.}~\bibnamefont{Bruno}},
  \bibinfo{author}{\bibfnamefont{V.~K.} \bibnamefont{Dugaev}},
  \bibnamefont{and}
  \bibinfo{author}{\bibfnamefont{M.}~\bibnamefont{Taillefumier}},
  \bibinfo{journal}{Phys. Rev. Lett.} \textbf{\bibinfo{volume}{93}},
  \bibinfo{pages}{096806} (\bibinfo{year}{2004}).

\bibitem[{\citenamefont{Tatara et~al.}(2007)\citenamefont{Tatara, Kohno,
  Shibata, Lemaho, and Lee}}]{Tatara_2007}
\bibinfo{author}{\bibfnamefont{G.}~\bibnamefont{Tatara}},
  \bibinfo{author}{\bibfnamefont{H.}~\bibnamefont{Kohno}},
  \bibinfo{author}{\bibfnamefont{J.}~\bibnamefont{Shibata}},
  \bibinfo{author}{\bibfnamefont{Y.}~\bibnamefont{Lemaho}}, \bibnamefont{and}
  \bibinfo{author}{\bibfnamefont{K.-J.} \bibnamefont{Lee}},
  \bibinfo{journal}{Journal of the Physical Society of Japan}
  \textbf{\bibinfo{volume}{76}}, \bibinfo{pages}{054707}
  (\bibinfo{year}{2007}).

\bibitem[{\citenamefont{Ye et~al.}(1999)\citenamefont{Ye, Kim, Millis,
  Shraiman, Majumdar, and Te\ifmmode \check{s}\else
  \v{s}\fi{}anovi\ifmmode~\acute{c}\else \'{c}\fi{}}}]{Ye1999}
\bibinfo{author}{\bibfnamefont{J.}~\bibnamefont{Ye}},
  \bibinfo{author}{\bibfnamefont{Y.~B.} \bibnamefont{Kim}},
  \bibinfo{author}{\bibfnamefont{A.~J.} \bibnamefont{Millis}},
  \bibinfo{author}{\bibfnamefont{B.~I.} \bibnamefont{Shraiman}},
  \bibinfo{author}{\bibfnamefont{P.}~\bibnamefont{Majumdar}}, \bibnamefont{and}
  \bibinfo{author}{\bibfnamefont{Z.}~\bibnamefont{Te\ifmmode \check{s}\else
  \v{s}\fi{}anovi\ifmmode~\acute{c}\else \'{c}\fi{}}}, \bibinfo{journal}{Phys.
  Rev. Lett.} \textbf{\bibinfo{volume}{83}}, \bibinfo{pages}{3737}
  (\bibinfo{year}{1999}).

\bibitem[{\citenamefont{Denisov et~al.}(2016)\citenamefont{Denisov, Rozhansky,
  Averkiev, and L\"ahderanta}}]{prl_skyrmion}
\bibinfo{author}{\bibfnamefont{K.~S.} \bibnamefont{Denisov}},
  \bibinfo{author}{\bibfnamefont{I.~V.} \bibnamefont{Rozhansky}},
  \bibinfo{author}{\bibfnamefont{N.~S.} \bibnamefont{Averkiev}},
  \bibnamefont{and}
  \bibinfo{author}{\bibfnamefont{E.}~\bibnamefont{L\"ahderanta}},
  \bibinfo{journal}{Phys. Rev. Lett.} \textbf{\bibinfo{volume}{117}},
  \bibinfo{pages}{027202} (\bibinfo{year}{2016}).

\bibitem[{\citenamefont{a and Kawamura}(2002)}]{Tatara}
\bibinfo{author}{\bibfnamefont{G.}~\bibnamefont{a}} \bibnamefont{and}
  \bibinfo{author}{\bibfnamefont{H.}~\bibnamefont{Kawamura}},
  \bibinfo{journal}{J. Phys. Soc. of Japan} \textbf{\bibinfo{volume}{71}},
  \bibinfo{pages}{2613} (\bibinfo{year}{2002}).

\bibitem[{\citenamefont{Onoda et~al.}(2004)\citenamefont{Onoda, Tatara, and
  Nagaosa}}]{Onoda_SkyrmNumber}
\bibinfo{author}{\bibfnamefont{M.}~\bibnamefont{Onoda}},
  \bibinfo{author}{\bibfnamefont{G.}~\bibnamefont{Tatara}}, \bibnamefont{and}
  \bibinfo{author}{\bibfnamefont{N.}~\bibnamefont{Nagaosa}},
  \bibinfo{journal}{Journal of the Physical Society of Japan}
  \textbf{\bibinfo{volume}{73}}, \bibinfo{pages}{2624} (\bibinfo{year}{2004}).

\bibitem[{\citenamefont{Ndiaye et~al.}(2017)\citenamefont{Ndiaye, Akosa, and
  Manchon}}]{Arab_Papa}
\bibinfo{author}{\bibfnamefont{P.~B.} \bibnamefont{Ndiaye}},
  \bibinfo{author}{\bibfnamefont{C.~A.} \bibnamefont{Akosa}}, \bibnamefont{and}
  \bibinfo{author}{\bibfnamefont{A.}~\bibnamefont{Manchon}},
  \bibinfo{journal}{Phys. Rev. B} \textbf{\bibinfo{volume}{95}},
  \bibinfo{pages}{064426} (\bibinfo{year}{2017}).

\bibitem[{\citenamefont{Ohe et~al.}(2007)\citenamefont{Ohe, Ohtsuki, and
  Kramer}}]{Simu_Chir}
\bibinfo{author}{\bibfnamefont{J.-i.} \bibnamefont{Ohe}},
  \bibinfo{author}{\bibfnamefont{T.}~\bibnamefont{Ohtsuki}}, \bibnamefont{and}
  \bibinfo{author}{\bibfnamefont{B.}~\bibnamefont{Kramer}},
  \bibinfo{journal}{Phys. Rev. B} \textbf{\bibinfo{volume}{75}},
  \bibinfo{pages}{245313} (\bibinfo{year}{2007}).

\bibitem[{\citenamefont{Metalidis}(2007)}]{Simulation}
\bibinfo{author}{\bibfnamefont{G.}~\bibnamefont{Metalidis}},
  \bibinfo{journal}{Dissertation} p.~\bibinfo{pages}{71}
  (\bibinfo{year}{2007}).

\bibitem[{\citenamefont{Berry}(1984)}]{berry1984quantal}
\bibinfo{author}{\bibfnamefont{M.~V.} \bibnamefont{Berry}}, in
  \emph{\bibinfo{booktitle}{Proceedings of the Royal Society of London A:
  Mathematical, Physical and Engineering Sciences}} (\bibinfo{organization}{The
  Royal Society}, \bibinfo{year}{1984}), vol. \bibinfo{volume}{392}, pp.
  \bibinfo{pages}{45--57}.

\bibitem[{\citenamefont{Lyanda-Geller et~al.}(2001)\citenamefont{Lyanda-Geller,
  Chun, Salamon, Goldbart, Han, Tomioka, Asamitsu, and Tokura}}]{Lyana-Geller1}
\bibinfo{author}{\bibfnamefont{Y.}~\bibnamefont{Lyanda-Geller}},
  \bibinfo{author}{\bibfnamefont{S.~H.} \bibnamefont{Chun}},
  \bibinfo{author}{\bibfnamefont{M.~B.} \bibnamefont{Salamon}},
  \bibinfo{author}{\bibfnamefont{P.~M.} \bibnamefont{Goldbart}},
  \bibinfo{author}{\bibfnamefont{P.~D.} \bibnamefont{Han}},
  \bibinfo{author}{\bibfnamefont{Y.}~\bibnamefont{Tomioka}},
  \bibinfo{author}{\bibfnamefont{A.}~\bibnamefont{Asamitsu}}, \bibnamefont{and}
  \bibinfo{author}{\bibfnamefont{Y.}~\bibnamefont{Tokura}},
  \bibinfo{journal}{Phys. Rev. B} \textbf{\bibinfo{volume}{63}},
  \bibinfo{pages}{184426} (\bibinfo{year}{2001}).

\bibitem[{\citenamefont{Sapozhnikov et~al.}(2016)\citenamefont{Sapozhnikov,
  Vdovichev, Ermolaeva, Gusev, Fraerman, Gusev, and Petrov}}]{Fraerman1}
\bibinfo{author}{\bibfnamefont{M.~V.} \bibnamefont{Sapozhnikov}},
  \bibinfo{author}{\bibfnamefont{S.~N.} \bibnamefont{Vdovichev}},
  \bibinfo{author}{\bibfnamefont{O.~L.} \bibnamefont{Ermolaeva}},
  \bibinfo{author}{\bibfnamefont{N.~S.} \bibnamefont{Gusev}},
  \bibinfo{author}{\bibfnamefont{A.~A.} \bibnamefont{Fraerman}},
  \bibinfo{author}{\bibfnamefont{S.~A.} \bibnamefont{Gusev}}, \bibnamefont{and}
  \bibinfo{author}{\bibfnamefont{Y.~V.} \bibnamefont{Petrov}},
  \bibinfo{journal}{Applied Physics Letters} \textbf{\bibinfo{volume}{109}},
  \bibinfo{eid}{042406} (\bibinfo{year}{2016}).

\bibitem[{\citenamefont{Landau and Lifshit︠s︡}(1977)}]{LL3}
\bibinfo{author}{\bibfnamefont{L.}~\bibnamefont{Landau}} \bibnamefont{and}
  \bibinfo{author}{\bibfnamefont{E.}~\bibnamefont{Lifshit︠s︡}},
  \emph{\bibinfo{title}{Quantum Mechanics: Non-relativistic Theory}},
  Butterworth-Heinemann (\bibinfo{publisher}{Butterworth-Heinemann},
  \bibinfo{year}{1977}).

\bibitem[{\citenamefont{Yin et~al.}(2015)\citenamefont{Yin, Liu, Barlas, Zang,
  and Lake}}]{Spin_Top_Hall}
\bibinfo{author}{\bibfnamefont{G.}~\bibnamefont{Yin}},
  \bibinfo{author}{\bibfnamefont{Y.}~\bibnamefont{Liu}},
  \bibinfo{author}{\bibfnamefont{Y.}~\bibnamefont{Barlas}},
  \bibinfo{author}{\bibfnamefont{J.}~\bibnamefont{Zang}}, \bibnamefont{and}
  \bibinfo{author}{\bibfnamefont{R.~K.} \bibnamefont{Lake}},
  \bibinfo{journal}{Phys. Rev. B} \textbf{\bibinfo{volume}{92}},
  \bibinfo{pages}{024411} (\bibinfo{year}{2015}).

\bibitem[{\citenamefont{Babikov}(1967)}]{Babikov1}
\bibinfo{author}{\bibfnamefont{V.~V.} \bibnamefont{Babikov}},
  \bibinfo{journal}{Usp. Fiz. Nauk} \textbf{\bibinfo{volume}{92}},
  \bibinfo{pages}{3} (\bibinfo{year}{1967}).

\bibitem[{\citenamefont{Adhikari}(1986)}]{Adhikari}
\bibinfo{author}{\bibfnamefont{S.~K.} \bibnamefont{Adhikari}},
  \bibinfo{journal}{American Journal of Physics} \textbf{\bibinfo{volume}{54}},
  \bibinfo{pages}{362} (\bibinfo{year}{1986}).

\end{thebibliography}

\end{document}